\documentclass[12pt]{iopart}
\newcommand{\vc}[1]{\mbox{\boldmath\/${#1}$}}
\usepackage{graphicx}
\usepackage{iopams,enumerate}

\begin{document}

\title[High order fluid model for streamer discharges. II. Investigation of planar fronts.]{High order fluid model for streamer discharges. II. Numerical solution and investigation of planar fronts}

\author{A.H. Markosyan$^{1}$, S. Dujko$^{1,2}$ and U. Ebert$^{1,3}$}
\address{$^1$ Centrum Wiskunde \& Informatica (CWI), P.O. Box 94079, 1090 GB Amsterdam, The Netherlands}
\address{$^2$ Institute of Physics, University of Belgrade, P.O. Box 68, 11080 Zemun, Belgrade, Serbia}
\address{$^3$ Department of Applied Physics, Eindhoven University of Technology, P.O. Box 513, 5600 MB Eindhoven, The Netherlands}

\ead{aram.markosyan@cwi.nl}

\begin{abstract}
The high order fluid model developed in the preceding paper is employed here to study the propagation of negative planar streamer fronts in pure nitrogen. The model consists of the balance equations for electron density, average electron velocity, average electron energy and average electron energy flux. These balance equations have been obtained as velocity moments of Boltzmann's equation and are here coupled to the Poisson equation for the space charge electric field. Here the results of simulations with the high order model, with a PIC/MC (Particle in cell/Monte Carlo) model and with the first order fluid model based on the hydrodynamic drift-diffusion approximation are presented and compared. The comparison with the MC model clearly validates our high order fluid model, thus supporting its correct theoretical derivation and numerical implementation. The results of the first order fluid model with local field approximation, as usually used for streamer discharges, show considerable deviations. Furthermore, we study the inaccuracies of simulation results caused by an inconsistent implementation of transport data into our high order fluid model. We also demonstrate the importance of the energy flux term in the high order model by comparing with results where this term is neglected. Finally, results with an approximation for the high order tensor in the energy flux equation is found to agree well with the PIC/MC results for reduced electric fields up to 1000 Townsend, as considered in this work.
\end{abstract}

\submitto{\JPD}
 
\section{Introduction}
\label{sec1}
Streamer discharges occur in nature and technology, predominantly in pulsed high voltage discharges; at their rapidly propagating tips electric fields well above the break-down value are maintained that create high electron energies and reaction rates. Streamers occur in lightning and sprites \cite{Pasko2006, EbertS2008, EbertNLLBV2010, LuqueE2010} as well as in industrial applications such as lighting \cite{Flesch2006, ListerLL2004, SobotaMVDH2010}, treatment of polluted gases and water \cite{Veldhuizen2000}, disinfection \cite{vanHeeschWP2008}, plasma jets and bullets \cite{Naidis2010, Naidis2011, LuLP2012, YousfiEMJ2012, BoeufYP2013} and plasma assisted combustion \cite{Starikovskaia2006, StarikovskiyA2013}. Further optimization and understanding of such applications depends on an accurate knowledge of the electron dynamics during streamer development.

A Monte Carlo technique is able to track the electron distribution in phase space, yielding both electron density profiles and electron energy distribution. However, a full particle model long has been computationally too demanding or too inaccurate due to the used super-particle techniques, and even now the parameter range of simulations is limited. For recent progress in particle and hybrid models we refer to~\cite{LiBEM2007, LiEH2010, LiEH2012, LuqueE2012, LiTNHE2012}. Due to the computational costs and limitations of particle models, up to today mainly fluid approximations have been used to model the structure and evolution of streamer discharges in two or three spatial dimensions. These fluid models were restricted almost exclusively to the hydrodynamic reaction drift diffusion approximation combined with the local field approximation, though the short-comings of this model have been acknowledged and documented \cite{LiBEM2007, LiEH2010, LiTNHE2012, LiEBH2008, LiEH2009, BayleC1985, GuoW1993, Naidis1997, KanzariYH1998, EichwaldDMYD2006, KorenEGGKK2012}. They arise from the fact that within the streamer front the electron density develops large spatial gradients and a complex interaction with the field. This in turn limits the use of the local field approximation as the electron energy does not immediately relax to the value determined by the local field, but depends upon the electric field in a wider spatial range.

The aim of the preceding and the present paper is to develop and test a better fluid approximation for streamer discharges in gaseous media. In the preceding paper~\cite{PaperI} we have derived a high order fluid model, whose predictions for streamer dynamics will be tested in the present paper. The high order fluid model contains equations for the electron density, for the average electron velocity, for the average electron energy and for the average electron energy flux. It was obtained from velocity moments of the Boltzmann equation and closed in the local mean energy approximation. Momentum transfer theory \cite{Robson1986, VrhovacP1996, WhiteRDNL2009} was used to evaluate the collisional terms in the balance equations, and particular emphasis was placed upon the correct representation of momentum and energy transfer in energy-dependent non-conservative collisions. The system was truncated at the level of the energy flux balance and simplifying approximations were introduced in the momentum and energy flux balance equations in order to close the system. In particular, in the momentum balance equation it was assumed that the distribution of velocities is isotropic assuming isotropy of the temperature and pressure tensors. Thus, the pressure tensor was reduced to a scalar kinetic pressure. Furthermore, the same approximation was used in the energy flux balance equation. If one assumes an isotropic distribution of velocities then the contribution of higher terms including the high order heat flux tensor and the high order pressure tensor is relatively small and could be neglected. However, the energy flux transported by the convective particle motion should be treated and implemented carefully. The high order tensor in the energy flux balance equation is expressed in terms of lower moments using the simplifying assumption that the pressure tensor is isotropic. Thus with such a treatment of the random motion and the high order terms associated with the energy flux of the drift motion, we have obtained a complete and closed system of fluid equations sufficient to determine all macroscopic streamer properties.

In the present paper the system of fluid equations derived in the preceding paper is solved numerically for a negative planar streamer ionization front and compared with results of a PIC/MC particle model and of the first order fluid model. The comparison with the particle model confirms that our high order fluid model approximates the particle behavior very well. The comparison with the commonly used first order fluid model of reaction drift diffusion type shows the short-comings of this classical model. This concerns the front velocities, the ionization levels behind the front as well as the electron energy distribution in the high field region ahead of the front and in the ionized low field region in the streamer interior. Actually the inaccurate ionization level in the streamer interior in the first order fluid model is a direct consequence of the inaccuracies of the electron energy distribution \cite{LiBEM2007,LiEH2010}. Inaccurate electron densities and energies also have direct consequences for the gas chemistry in the streamer interior~\cite{LuqueGV2011, FlittiP2009, Pancheshnyi2005, Popov2001, Popov2003}. In addition to the accuracy and comprehensiveness of our high order fluid model, its firm theoretical foundation is certainly an additional favoring factor.

We begin by presenting in section \ref{sec2} the differential equations governing the electron density, the average electron velocity, the average electron energy and the average electron energy flux. Then we discuss the numerical algorithm used for the solution of the differential equations. The specific elements of this discussion include a brief mathematical reminder on hyperbolic systems, the implementation of initial and boundary conditions and the discretization in space and time. Our numerical analysis is performed in a 1D model, because in the present paper we are not interested in a streamer morphology and related phenomena where a full multidimensional model is required. We use a PIC/MC method as an alternative technique to verify our high order fluid model. Following the recent work of Li {\it et al.} \cite{LiBEM2007,LiEH2010,LiEH2012}, in section \ref{sec3} we give a brief description of this method with particular emphasis upon the construction of planar fronts in the particle model. The high order fluid model is then applied in section \ref{sec4} where examples of planar fronts in pure nitrogen are presented. The effects of photoionization are not considered and present theory can be applied to negative fronts in discharges where photoionization is very weak (high-purity nitrogen is a good example). Results demonstrating the validity of a high order fluid model compared with the PIC/MC are shown. Following the preceding paper, we demonstrate the importance of a consistent implementation of transport data in fluid models, but now instead of using the first order model as in the preceding paper, we employ our high order model. Along similar lines, the accuracy of the two term approximation for solving Boltzmann's equation in the context of high order fluid studies of the streamer discharges is examined. We pay particular attention to the role of the electron energy flux. For illustrative purposes, calculated results for the planar fronts based on a high order model with energy flux and those calculated without energy flux, are compared over a range of electric fields in order to verify the physical arguments associated with the solution regimes and closure assumptions outlined in the preceding paper. Next, we examine the closure assumption associated with the explicit influence of the high order tensor appearing in the energy flux equation on the streamer dynamics. We conclude that the explicit contribution of the high order tensor in the energy flux equation is negligible and it makes a notable difference only in the velocity of the planar fronts. In summary, in this paper we present the numerical solution of a high order fluid model, test and verify it on the more microscopic PIC/MC model, and discuss how inherent streamer properties deviate when simpler fluid models are used.
\section{Model description and numerical solution}
\label{sec2}

\subsection{The high order fluid model}

In our preceding paper~\cite{PaperI} we have developed a high order fluid model for streamer discharges. The balance equations were obtained as velocity moments of the Boltzmann equation while the collisional terms were evaluated using momentum transfer theory. We have truncated the system of fluid equations at the level of energy flux balance by reducing the pressure tensor to a scalar kinetic pressure and by neglecting the high order terms associated with the flux of thermal motion. The energy flux of the drift motion, however, is included. For more details of the derivation we refer to~\cite{PaperI}. The derived model consists of a set of differential equations for the electron density $n$, for the average electron velocity $\vc{v}$, for the average electron energy $\varepsilon$ and for the average electron energy flux $\vc{\xi}$:
\begin{equation}
\frac{\partial n}{\partial t} + \nabla \cdot n\vc{v} = -n\big(\widetilde{\nu}_A - \widetilde{\nu}_I\big)\,,
\label{2.1}
\end{equation}
\begin{equation}
\frac{\partial}{\partial t} (n\vc{v}) + \frac{2}{3m}\nabla(n\varepsilon)-n\frac{e}{m}\vc{E} = -n\vc{v}\Big(\widetilde{\nu}_m + \widetilde{\nu}_I + \frac{2}{3m}\varepsilon\widetilde{\nu}_a^{\,'}\Big)\;,
\label{2.2}
\end{equation}
\begin{eqnarray}
\frac{\partial}{\partial t}(n\varepsilon)+&\nabla\cdot(n\vc{\xi})-e\vc{E}\cdot(n\vc{v}) = -n\widetilde{\nu}_e\left(\varepsilon-\frac{3}{2}kT_0 \right)\nonumber \\
&-n\sum_{i}(\widetilde{\nu}_i-\widetilde{\nu}_i^{\;s})\epsilon_i-
n\varepsilon\widetilde{\nu}_A-n\sum_i\widetilde{\nu}_i^{(i)}\Delta\epsilon_i^{(i)}\;,
\label{2.3}
\end{eqnarray}
\begin{equation}
\frac{\partial}{\partial t}(n\vc{\xi}) + \nabla\Big(\beta\frac{2n}{3m}\varepsilon^2 \Big)-\frac{5}{3}n\varepsilon e\vc{E} = -\widetilde{\nu}_mn\vc{\xi}\,,
\label{2.4}
\end{equation}
where $\vc{E}$ is the electric field, $m$ and $e$ are electron mass and charge, $T_0$ is gas temperature and $k$ is the Boltzmann constant. The collision frequencies for momentum $\widetilde{\nu}_m$ and energy $\widetilde{\nu}_e$ transfer are given by equations (49) and (50) of the preceeding paper while $\widetilde{\nu}_I$ and $\widetilde{\nu}_A$ are the ionization and attachment rate coefficients. $\widetilde{\nu}_i$ and $\widetilde{\nu}_i^{\,s}$ are inelastic and superelastic collision frequencies, respectively for inelastic channel $i$. $\beta$ is a parameter introduced to approximate the high order tensors in the energy flux equation in terms of lower moments (see equation (55) of the preceeding paper).

Taking into account the rapid growth and propagation of the streamers with orders of the electron drift velocity and the relatively slow drift velocity of ions, the ions are approximated as immobile, as usually done when streamer ionization fronts are analyzed. Therefore, the charge densities due to the positive and negative ions change only due to ionization and attachment:
\begin{equation}
\label{2.5}
\frac{\partial n_{ion}}{\partial t} = -n\big(\widetilde{\nu}_A - \widetilde{\nu}_I\big)\,.
\end{equation}
In order to account for the space charge effects, the system (\ref{2.1})-(\ref{2.5}) is coupled with the Poisson equation for the potential $\phi$ which then enables the calculation of the electric field $\vc E$:
\begin{equation}
\nabla^2 \phi = -\,\frac{e}{\epsilon_0}(n_{ion} - n), \quad \vc{E} = -\nabla\phi\,,
\label{2.6}
\end{equation}
where $\epsilon_0$ is the dielectric constant.

\subsection{1D hyperbolic system of balance laws}
\label{sec2.1}
In the present paper we simulate the propagation of negative streamer fronts in pure $\textrm N_2$ at atmospheric pressure and at an ambient temperature of $298$ \textrm K in 1D. As $\textrm N_2$ is a non-attaching gas, electron attachment does not appear in the source terms of our system (\ref{2.1})-(\ref{2.6}). The effects of superelastic collisions are not taken into account and we consider only single ionization with ionization energy $\epsilon_I$. The electron dynamics of equations (\ref{2.1})--(\ref{2.4}) in 1D is given by the following nonlinear system of balance laws:
\begin{equation}
\label{2.7}
\frac{\partial \vc{u}}{\partial t}+\vc{ A}(\vc{ u})\frac{\partial \vc{ u}}{\partial x}=\vc{ F}(\vc{u}),
\end{equation}
where the primitive variables are
\begin{equation}
\vc{u}=(n,nv,n\varepsilon,n\xi)^\mathbf T,
\label{2.8}
\end{equation}
the matrix $\vc{ A}(\vc{ u})$ is
\begin{equation}
\vc{ A}(\vc{ u})=\left(\begin{array}{cccc}0 & 1 & 0 & 0 \\0 & 0 & \frac{2}{3m} & 0 \\0 & 0 & 0 & 1 \\-\beta \frac{2\varepsilon ^2}{3m} & 0 & \beta \frac{4\varepsilon}{3m} & 0\end{array}\right),
\label{2.9}
\end{equation}
and the source term is
\begin{equation}
\vc{ F} (\vc{u})  = \left(\begin{array}{c}n\widetilde{\nu}_I \\\frac{nqE}{m}-nv\left(\widetilde{\nu}_m+\widetilde{\nu}_I\right) \\qEnv-n\Big[ \widetilde{\nu}_e\left( \varepsilon - \frac{3}{2}kT_0\right) +\sum\limits_{i} {\widetilde{\nu}}_{i}
                           \epsilon_{i}+ \widetilde{\nu}_I\epsilon_I\Big] \\\frac{5qE}{3m}n\varepsilon -n\xi \widetilde{\nu}_m\end{array}\right)\,.
\label{2.10}
\end{equation}
The equation for the ions (\ref{2.5}) is without electron attachment
\begin{equation}
\label{2.11}
\frac{\partial n_{ion}}{\partial t} = n \widetilde{\nu}_I\,,
\end{equation}
while the Poisson equation (\ref{2.6}) in 1D has the following simple form:
\begin{equation}
\label{2.12}
\frac{\partial E}{\partial x} = \frac{e}{\epsilon_0}(n_{ion} - n)\,.
\end{equation}
Hence, the streamer dynamics in 1D for a non-attaching gas is described by the system of equations (\ref{2.7})-(\ref{2.12}).

Now we can apply the theory for hyperbolic systems of balance laws \cite{Leveq1, Leveq2, MortonM} to our system (\ref{2.7})-(\ref{2.9}). The matrix $\vc A(\vc u)$ has four eigenvalues:
\begin{equation}
\begin{array}{c}\lambda_{1,2} =  \pm\gamma \sqrt{\beta-\sqrt{\beta(\beta-1)}} \\\ \lambda_{3,4} = \pm\gamma \sqrt{\beta+\sqrt{\beta(\beta-1)}} \end{array},
\label{2.13}
\end{equation}
where $\gamma=\sqrt{\frac{2\epsilon}{3m}}$. All eigenvalues are real and distinct when
\begin{equation}
\label{2.14}
\beta = 0 \quad \textrm{or} \quad \beta\geq 1.
\end{equation}
This means that the system (\ref{2.7}) is \textit{hyperbolic} if the condition (\ref{2.14}) holds. Although the eigenvalues (\ref{2.13}) are simple, the corresponding right and left eigenvectors have a rather complicated form, which makes it impossible to work with the vector of Riemann invariants.

\subsection{Initial and boundary conditions}
\label{sec2.2}
The electric field $\vc{E}=E \hat{\vc e}_x$ (where $\hat{\vc e}_x$ is the unit vector in the $x$ direction) drives the dynamics. We take $E$ as a positive value; therefore electrons drift to the left, and negative streamer ionization fronts move to the left as well. To create steady propagation conditions for the negative front, the electric field on the left boundary $x=0$ is fixed to the time independent value $E_0$
\begin{equation}
E(0,t)= E_0 >0.
\label{2.15}
\end{equation}
The electric field for $x>0$ is calculated by integrating equation (\ref{2.12}) numerically over $x$, with (\ref{2.15}) as a boundary condition. The right boundary is located at $x=L$; in all calculations we set the system length $L$ to 1.2~mm.

All simulations are started with the same initial Gaussian type distribution for electrons and ions
\begin{equation}
n(x)|_{t=0}=n_i\exp{\Bigg[-\frac{(x-x_0)^2}{\sigma^2}\Bigg]}\,,
\label{2.14b}
\end{equation}
where we have chosen $n_i=2\times10^{18}$~m$^{-3}$, $x_0=0.8$~mm and $\sigma=0.029$~mm. The initial conditions for the average electron velocity, average electron energy and average electron energy flux are taken to be spatially homogeneous. The actual values of these quantities are calculated using a multi term solution of Boltzmann's equation, as already discussed in the preceding paper.

As discussed in the appendix, we use homogeneous Neumann boundary conditions for all components of $\vc u$ at both ends of the system, so that all electron quantities may flow out of the system. Only for the electron density at $x=L$ we employ a homogeneous Dirichlet boundary condition to prevent electrons from diffusing out. However, all calculations end before the ionization reaches the boundary, that means, before the boundary conditions start to become relevant.

Further details on the numerical discretization of the system can be found in the appendix.
\section{MC particle model and first order fluid model}
\label{sec3}

In the next section, we will compare the simulation results of our high order model with those of the PIC/MC particle model and with the first order fluid model. Here these models are briefly described.

\subsection{The MC particle model}

A Monte Carlo model following the motion of individual electrons contains the full physics that is to be approximated by a fluid model. The successful comparison of a fluid model with such a particle model validates the fluid model; of course, the same cross-sections have to be used in both models.

The construction of a planar front is straight forward in fluid models, as the spatial derivatives are simply evaluated in one direction. However, in the particle model electrons move in all three spatial dimensions and hence, the three-dimensional setting has to be restricted as in previous work \cite{LiBEM2007}. An essentially one-dimensional setting is achieved by considering only a small transversal area $\vc T$ of the front and by imposing periodic boundary conditions at the lateral boundaries. The electric field is calculated only in the forward direction $x$ through
\begin{equation}
\label{mc-el}
E_x(x,t) = E_x(x_0,t)+\int^x_{x_0}dx'\int_{\vc T} \frac{dy\ dz}{\vc T}\frac{e(n_{ion}-n)(x',y,z,t)}{\epsilon_0}\,,
\label{2.28}
\end{equation}
where $E_x$ is the electric field in the $x$-direction, and $x_0$ is an arbitrary position. Therefore, fluctuations of the transversal field due to density fluctuations in the transversal direction are not included. The density fluctuations projected onto the forward direction depend on the transversal area $\vc T$ over which the averages are taken.

\subsection{The first order fluid model}

The first order fluid model is the multiply used reaction drift diffusion model that also was called the ``classical fluid model'' in our previous work~\cite{LiEH2010, LiTNHE2012}; in these previous papers, it was already shown that the model can only serve as a first approximation, but cannot reproduce the results of a particle simulation quantitatively.

The steps of derivation of this model were discussed and criticized in our previous paper~\cite{PaperI}. The model is based on the hydrodynamic reaction drift diffusion approximation
\begin{equation}
\frac{\partial n}{\partial t} - \frac{\partial}{\partial x}\left( n\mu E + D\frac{\partial n}{\partial x}\right)=n\nu_I\,,
\label{2.27}
\end{equation}
where $\mu$ and $D$ are electron mobility and diffusion coefficient, respectively. Transport coefficients used as an input in this model are functions of the local electric field as discussed in the preceding paper. In order to account for the space charge effects, equation (\ref{2.27}) is coupled to the equations for ion density (\ref{2.11}) and for the electric field (\ref{2.12}). For the purpose of comparison, we use the same numerical schemes to discretize equation (\ref{2.27}) and the same initial and boundary conditions as introduced above for the high order model.

\section{Results and discussion}
\label{sec4}

In this section we present and compare the simulations results obtained with our high order fluid model, with the PIC/MC method and with the first order fluid model. We consider the configuration described in Section \ref{sec2.2}. The transport coefficients required as an input in both the first and the high order fluid model are given in our preceding paper~\cite{PaperI}. They were calculated from cross sections for electron scattering in N$_2$ through a multi term solution of the Boltzmann equation. Details of the calculation together with the prescription how to use the data in modeling are given in the same article. It is important to emphasize that the same collisional cross sections for electrons in N$_2$ are used in all models presented in this paper.

\subsection{Simulation of planar fronts: overall comparison between the models}
\label{sec4.1}

\begin{figure} [!htb]
\centering
\includegraphics[scale=0.3]{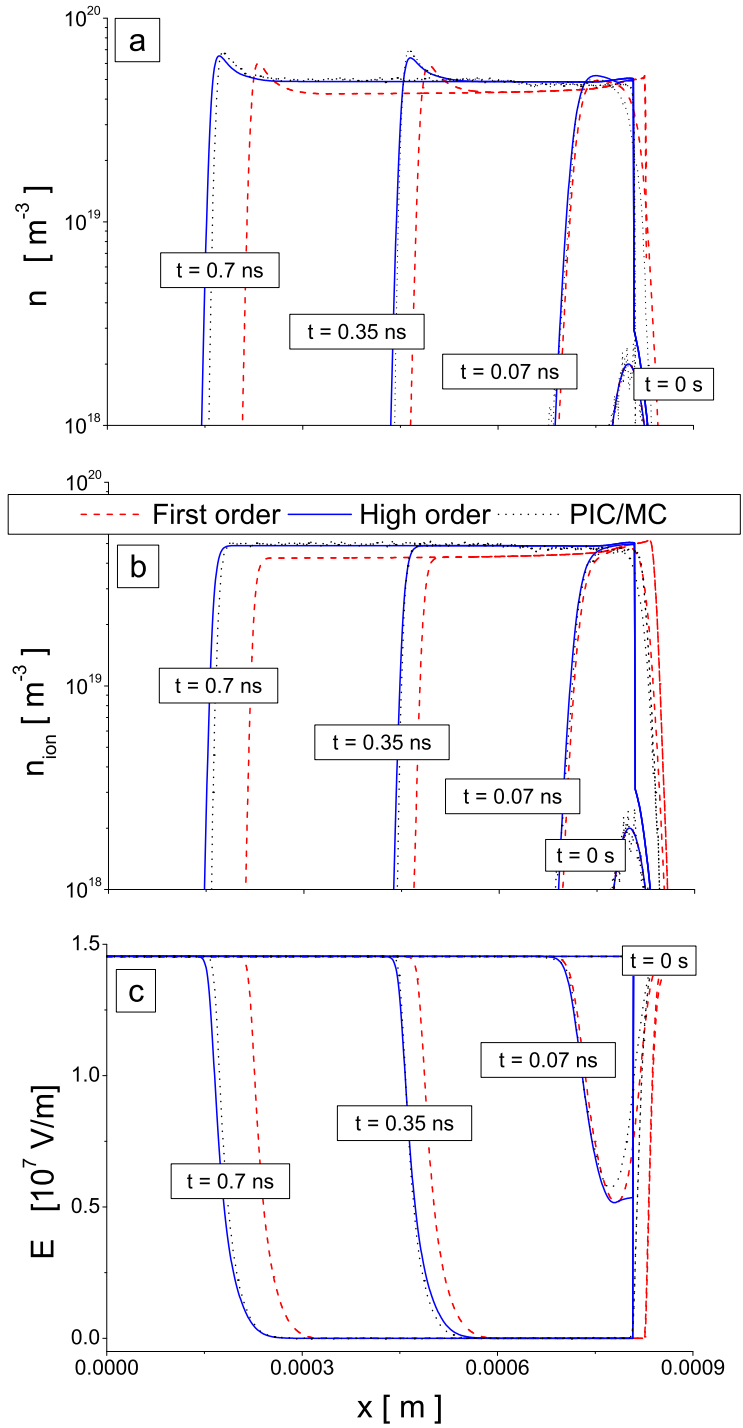}
\caption{Temporal evolution of the electron density (a), ion density (b) and electric field (c) in a planar front in N$_2$. Shown are the spatial profiles obtained with three different models as indicated on the graph. Flux transport data are used as an input in fluid models. The externally applied reduced electric field is 590 Td.}
\label{Fig3.1.1}
\end{figure}

Figure \ref{Fig3.1.1} shows the temporal evolution of the electron and ion densities and of the electric field in N$_2$, when the electric field ahead of the front is 590 Td (or equivalently 145 kV/cm at atmospheric pressure and at a temperature of 298 K). The initially Gaussian electron density shows the behavior observed many times in past: first, it grows due to the ionization processes; then charge separation occurs in the electric field due to the drift of oppositely charged particles in opposite directions, and the initially homogeneous electric field is distorted; and finally, when the field in the ionized region becomes more and more screened, the ionization stops in the ionized region and the typical ionization front profiles of electron and ion densities and of the electric field are established. The initial ionization avalanche is then said to have developed into a streamer.

\begin{figure} [!htb]
\centering
\includegraphics[scale=0.35]{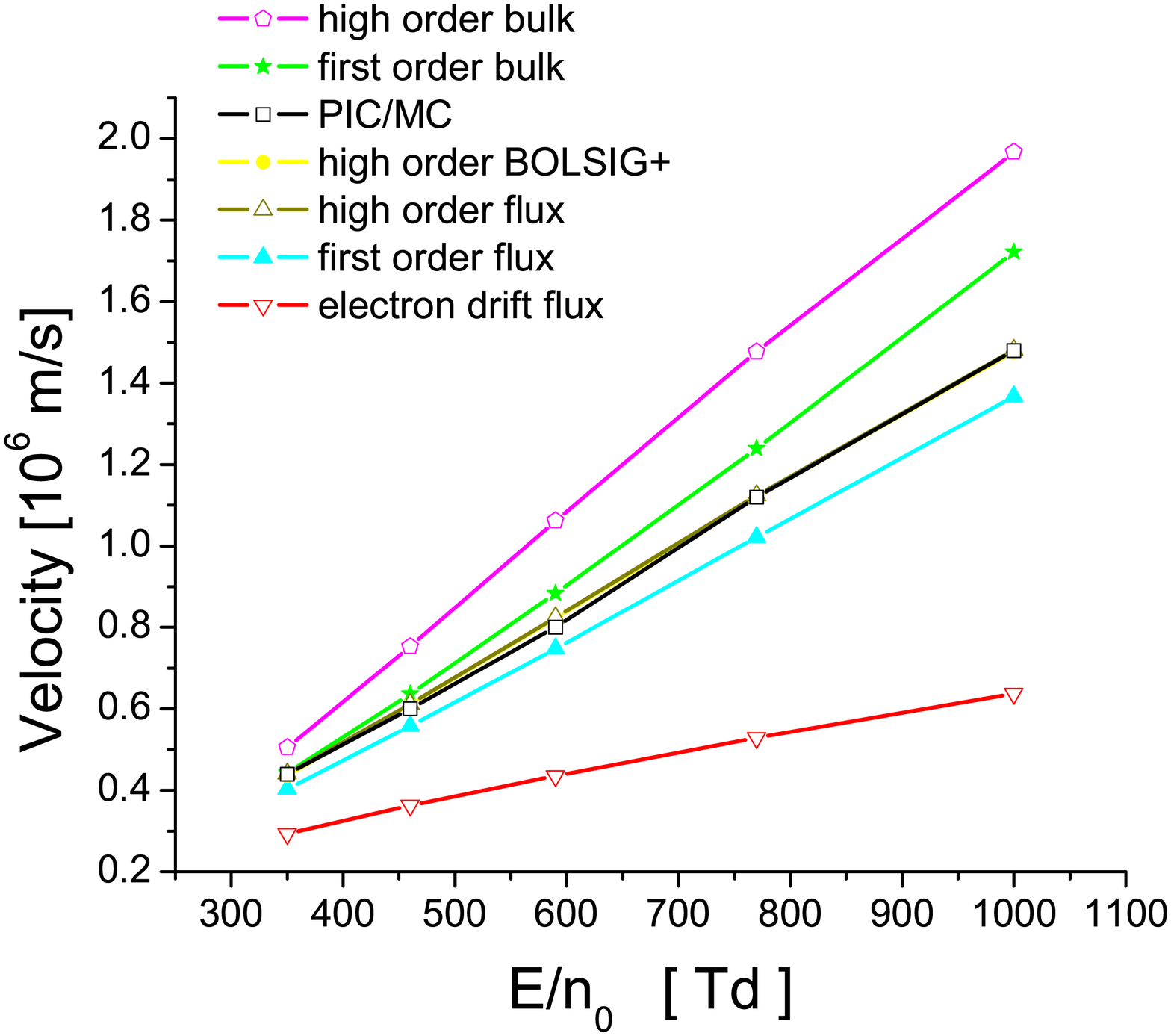}
\caption{Velocities of planar fronts as a function of the electric field obtained with the first and the high order fluid model and with the PIC/MC method; different sets of input data are used as indicated on the graph. The electron flux drift velocity as a function of $E/n_0$ is also included. We remark that the negative sign of all velocities is removed here.}
\label{Fig3.1.4}
\end{figure}

We note that the results of the high order fluid model and of the PIC/MC simulation agree very well. The streamer velocity is almost the same, as well as the electric field and the electron/ion density. Conversely, when the first order fluid model is applied, the streamer velocity is lower, with a lower electron density in the streamer head and in the streamer channel. In figure \ref{Fig3.1.4} we show how the streamer velocity depends on the applied reduced electric field $E/n_0$. In order to calculate the streamer velocity we have followed the evolution of a certain level ($2n_i$) of the electron density at the streamer front. We see that for increasing $E/n_0$ the differences of the streamer velocities between the first order and the high order fluid model increase slightly. On the other hand, the velocities of the high order fluid model and the PIC/MC agree very well. Similar trends are observed for the electron density behind the ionization front. While the first order model generally underestimates this electron density, the high order model approximates the PIC/MC results much better. This agreement holds in the full range of fields $E/n_0$ explored, while the differences between first order model and MC model increase with increasing $E/n_0$.

In some aspects the results presented in figures \ref{Fig3.1.1} and \ref{Fig3.1.4} are consistent with observations by Li {\it et al.} \cite{LiBEM2007, LiEH2010} and by Kanzari {\it et al.} \cite{KanzariYH1998}. In particular, Li {\it et al.} \cite{LiBEM2007, LiEH2010, LiTNHE2012} already have found considerable differences in velocities and ionization densities between MC particle model and classical fluid model; they have proposed an extension of the first order fluid model in~\cite{LiEH2010}, based on a phenomenological gradient expansion suggested in~\cite{Aleksandrov1996, Naidis1997b} to get a better agreement of the fluid model with the results of the particle model. The extended fluid model of Li and co-workers uses a density gradient expansion of the source term to approximate the spatial non-locality of the ionization processes at the streamer front. However, this model overestimates the particle results for the electron density in the streamer channel when the field is below 125 kV/cm (for standard temperature and pressure of the background gas) by up to 4~\%, while for an electric field of up to 200~kV/cm, the extended fluid model underestimates the particle results by up to 9~\%~\cite{LiEH2010}. On the other hand, the differences in the electron density behind the ionization front between our systematically derived high order fluid model and the PIC/MC model do not exceed 5\% for electric fields of up to 245~kV/cm as considered in this work, thus demonstrating that we correctly derived and implemented our high order fluid model.

\subsection{Electron energy profiles in different front regions within the different models}

\begin{figure} [!htb]
\centering
\includegraphics[scale=0.35]{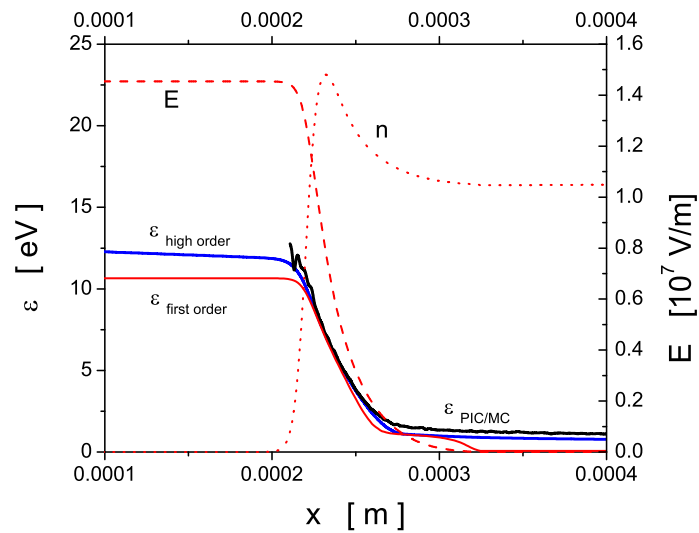}
\caption{The average electron energy in three different models at time 0.7 ns. Profiles of the electric field and the electron number density from the first order fluid model are also included. Calculations are performed for $E/n_0$ of 590 Td.}
\label{Fig3.1.5}
\end{figure}

We now focus on the characteristic differences in the profiles of the electron density. Figure \ref{Fig3.1.5} shows the profiles of the average electron energy for all three models, as well as the profiles of the electric field and of the electron density at $t=0.7$ ns for the first order model only. In the first order model, the average electron energy is assumed to adapt to the electric field instantaneously, therefore it was derived from the local electric field for the plot. However, the profile of the average energy obtained by the high order fluid and PIC/MC models is more complex.

We distinguish three different spatial regimes, the streamer head where most ionization occurs and where electron density and electric field have large gradients, the region ahead of the front where the electric field is high and the electron density vanishes, and the streamer interior where the electron density is finite and the electric field is low or vanishing.

First, figure \ref{Fig3.1.5} shows that in the streamer head the average electron energy given by the high order fluid and PIC/MC methods are higher than in the first order fluid model. Since the ionization rate in this energy range is an increasing function of the electron energy, it is clear that the higher the average electron energy, the higher is the ionization rate. These observations are consistent with those made by Li {\it et al.} \cite{LiBEM2007,LiEH2010}.

Second, in the region ahead of the front the average energy in the high order model has a slope which reflects the so-called non-local effects. In this region, the PIC/MC method is not efficient as there are not enough particles to sample the spatially resolved average electron energy correctly. It must be emphasized that the spatial variation of the average electron energy is present even in the  avalanche phase where an initial spatially homogeneous electric field is not distorted due to space charge effects. According to arguments given in~\cite{LiBEM2007}, the leading edge of an ionization front has essentially the same dynamics and therefore also the same energy slope as an electron swarm. In the past, many swarm oriented studies have been performed to explore the effects of spatial variation of the average energy through the swarm and many phenomena have been discovered \cite{WhiteRDNL2009,DujkoWRP2012,PetrovicRDM2002,WhiteNR2002}. Almost certainly, one of the most striking phenomena is the difference between the flux and the bulk transport data which follows directly from spatially dependent nonconservative collisions (ionization and/or electron attachment) resulting from a spatial variation of average electron energies within the swarm \cite{WhiteRDNL2009,DujkoWRP2012,DujkoEWP2011,NessR1986}.

Third, figure \ref{Fig3.1.5} shows strong disagreement between the average energy in the first order model and the average energies predicted by high order fluid and PIC/MC methods in the streamer interior. Within the first order fluid model the average electron energy attains approximately the thermal value. This follows directly from the local field approximation and from our Boltzmann equation calculation of the mean energy in the limit of vanishing electric field. It should be emphasized that thermal effects of background molecules are included in our Boltzmann equation analysis. In contrast to the first order fluid model, the average electron energy given by high order fluid and PIC/MC methods are significantly higher. This is another manifestation of the non-local effects, but this time, these phenomena take place in the streamer channel. As the velocity of the front is higher than the electron drift velocity even in the streamer head region, electrons slowly relax to lower energies in the lower or vanishing field in the streamer interior. This causes the spatial decay of average electron energy in high order fluid and PIC/MC model, that can be observed in figure \ref{Fig3.1.5}. This decrease of the average electron energy follows directly from the energy dependence of the collision frequency for energy relaxation shown in figure 2 of the preceding paper~\cite{PaperI}.
\begin{figure} [!htb]
\centering
\includegraphics[scale=0.35]{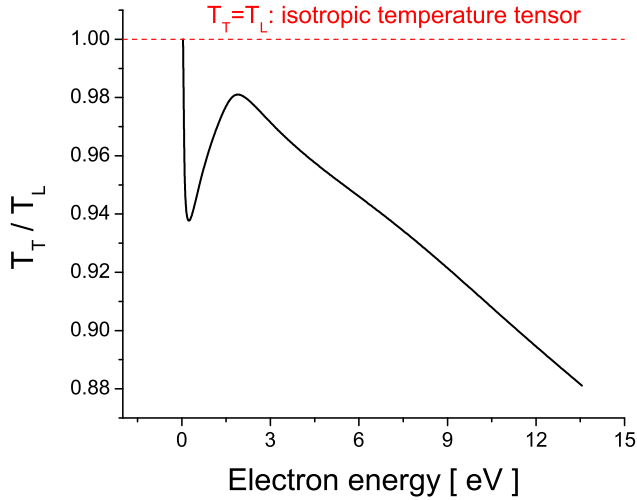}
\caption{The ratio between transverse and longitudinal components of the temperature tensor as a function of electron energy in N$_2$ based on the multi term approach for the Boltzmann equation.}
\label{Fig3.1.4b}
\end{figure}

Another interesting feature is the disagreement between the high order and the PIC/MC results for the average electron energy in the streamer channel. We believe that this disagreement arises from the limitations associated mainly with the high order fluid model. Although our PIC/MC method does not treat thermal effects of the background molecules, electron energies are too high for this to be significant. Most probably this disagreement arises from an increased anisotropy of the distribution function which in turn limits the adequacy of the closure assumptions associated with the pressure tensor. Indeed, the average energy in the streamer channel varies between 0.5 and 1 eV and exactly in the same energy region the cross sections for vibrational excitation grow rapidly in magnitude (a few orders of magnitude in a very narrow energy range) while the cross section for momentum transfer in elastic collisions varies relatively slowly with the electron energy. This will induce strong anisotropy of the distribution function, and it is clear that the assumption of the isotropy of the temperature tensor is certainly problematic under these conditions. These physical arguments are verified on figure \ref{Fig3.1.4b} where the ratio between transverse and longitudinal components of the temperature tensor for electrons in N$_2$ is shown. It shows a deep minimum in the profile between 0.5 and 1 eV which is a clear sign of an increasing anisotropy of the distribution function in velocity space. It should be emphasized that for the range of electric fields considered in this work, the different average energies in the streamer channel observed in high order fluid or PIC/MC model cannot induce significant changes in the electron density. Another physical argument which can be used to address the differences between the average electron energies in the streamer channel is associated with the accuracy of the PIC/MC method. In the energy range around 1 eV one must carefully follow the electrons in a PIC/MC simulation. If the time step between two successive collisions is too big then the energy losses due to rapidly increasing cross sections for vibrational excitation are not going to be included accurately. This is of less importance for the collisional processes whose cross sections do not vary so rapidly with the electron energy. One way to overcome this problem is to reduce the time step in a PIC/MC method but the penalty is a significant increase in the computation time. It is clear that additional testing and calculations are required to resolve this issue.

\begin{figure} [!htb]
\centering
\includegraphics[scale=0.35]{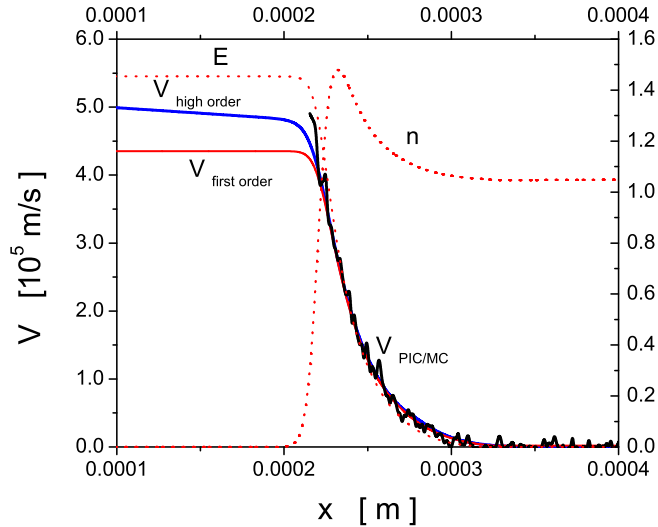}
\caption{Profiles of the average electron velocity in the first and the high order fluid models and of the PIC/MC model at time 0.7 ns. Profiles of the electric field and of the electron number density from the first order model are also included. Calculations are performed for $E/n_0$ of 590 Td. We remark that all velocities are negative, and their absolute value has been plotted.}
\label{Fig3.1.6}
\end{figure}

Figure \ref{Fig3.1.6} shows the average electron velocity through the streamer front. As for the average electron energy, in the first order model the average velocity of electrons is obtained from the profile of the electric field assuming the local field approximation. We see that the average velocity of electrons follows changes in the electric field instantaneously: in the outer streamer region where the electric field is constant the average velocity has also a constant value while in the streamer channel the average velocity is essentially zero. When considering the profiles of the average velocity derived by the high order model, one may observe the characteristic slope of this quantity in the outer region of the streamer while in the streamer channel it vanishes. This stands in contrast to the profile of the average energy. The relaxation of the average velocity is determined by the momentum transfer collision frequency which is a factor of $2.5\times10^4$ larger than the collision frequency for energy transfer. This means that the relaxation of the average velocity is much faster and hence in contrast to the average electron energy this quantity relaxes very rapidly. Figure \ref{Fig3.1.6} shows a good agreement between results derived by fluid models and by the PIC/MC method. As in the case of the average electron energy, the PIC/MC method is not suitable for the determination of the average velocity of electrons in the region ahead of the front because in this region only a few electrons exist.

\begin{figure} [!htb]
\centering
a) \includegraphics[scale=0.3]{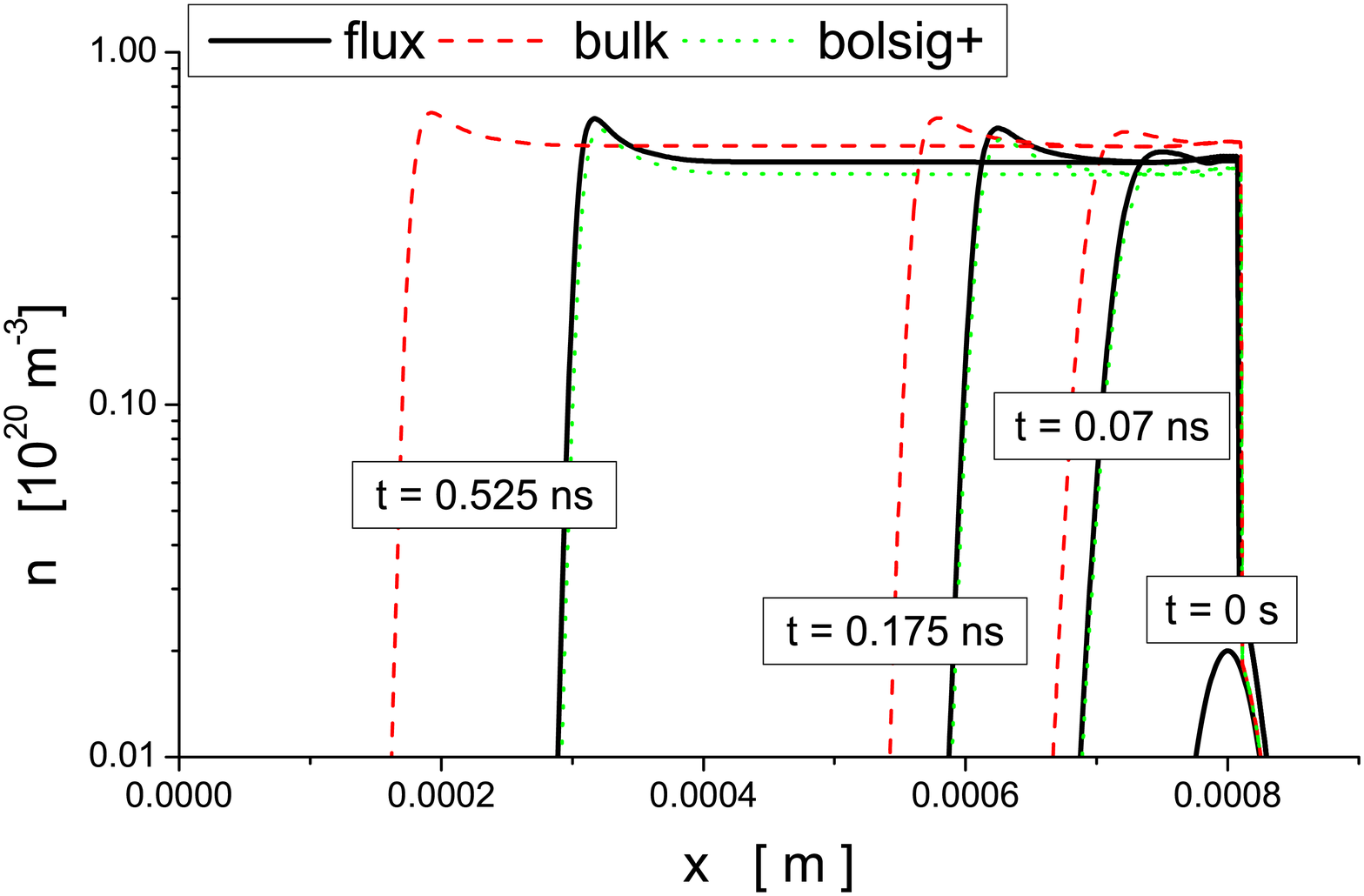}\\
\label{Fig3.2.1}
b) \includegraphics[scale=0.3]{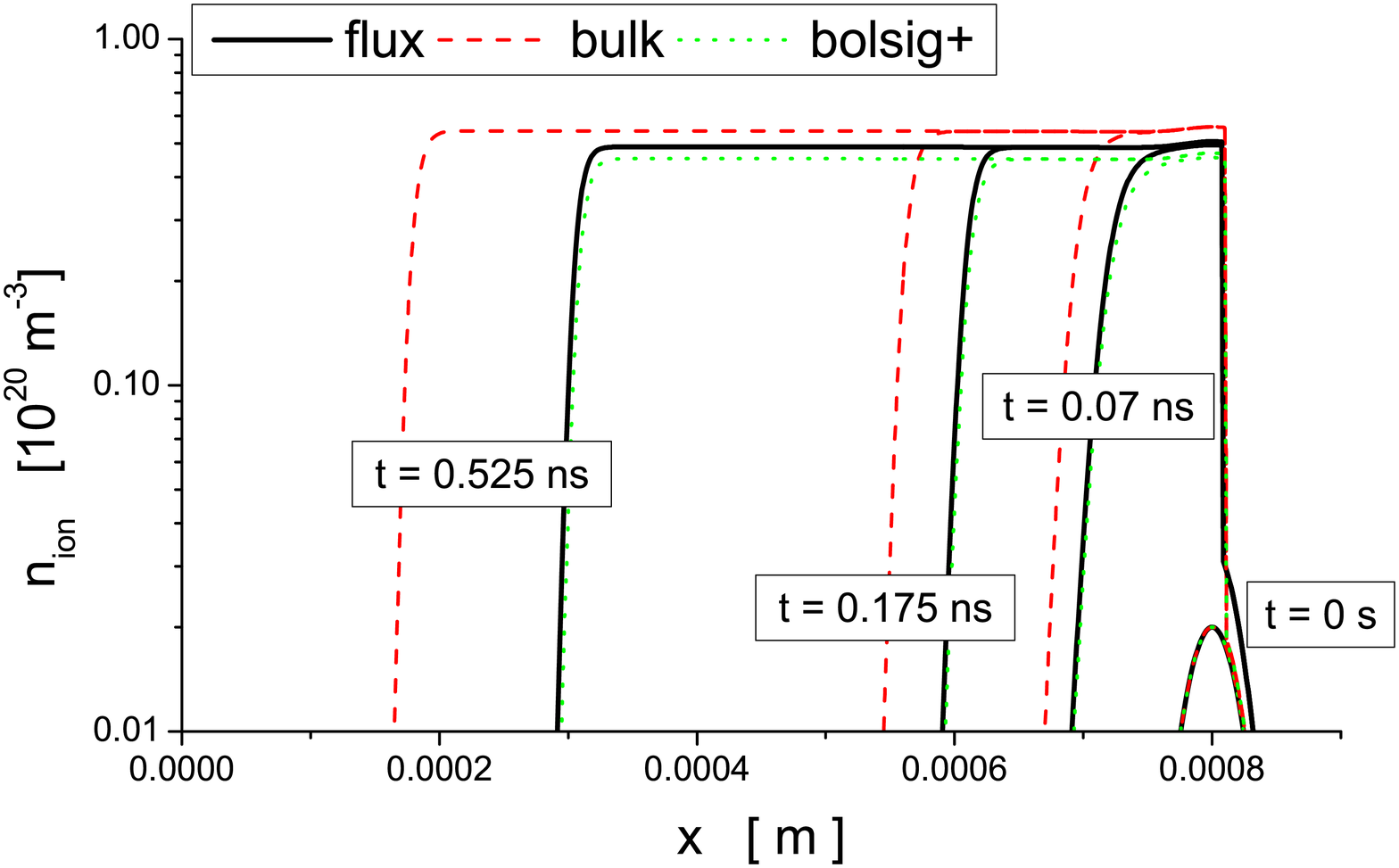}\\
\label{Fig3.2.2}
c) \includegraphics[scale=0.3]{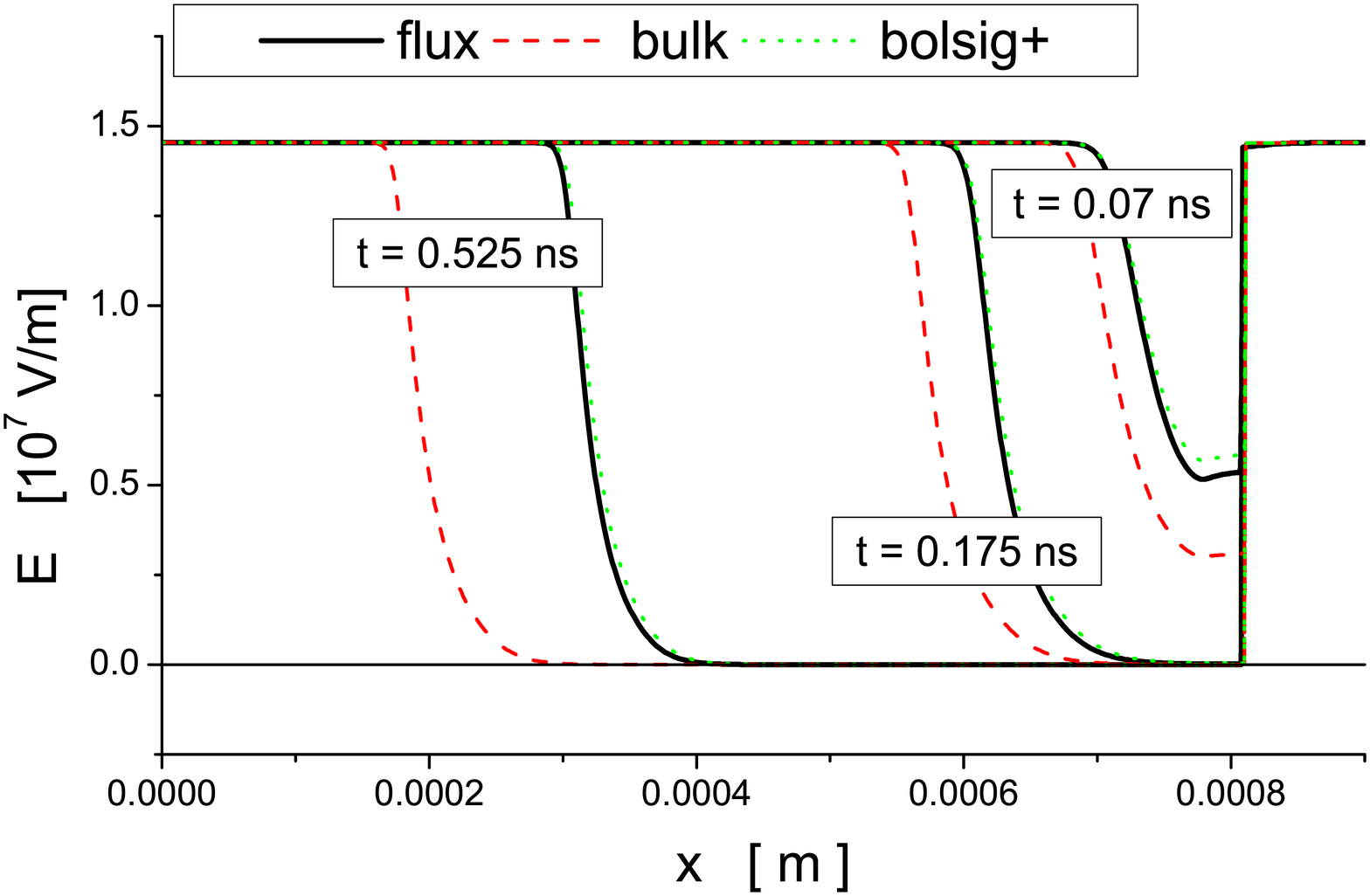}
\label{Fig3.2.3}
\caption{Temporal evolution of (a) the electron density, (b) the ion density and (c) the electric field in a planar negative ionization front in N$_2$ with a reduced electric field $E/n_0$ of 590 Td ahead of the front. Displayed are the spatial profiles obtained with three different sets of input data as indicated in the graph.}
\label{fig6}
\end{figure}
\begin{figure} [!htb]

a) \includegraphics[scale=0.3]{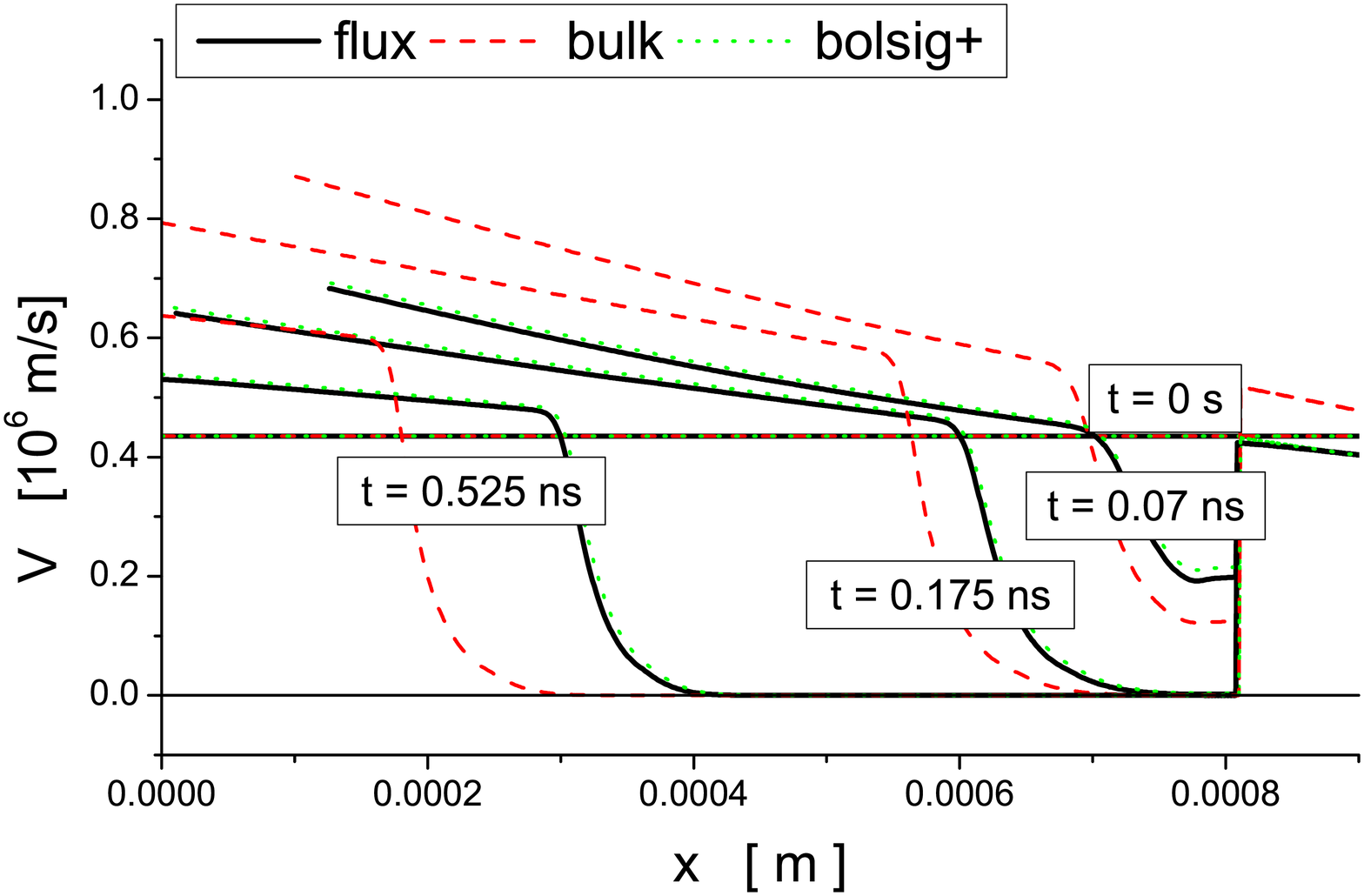}\\
\label{Fig3.2.6}
\centering
b) \includegraphics[scale=0.3]{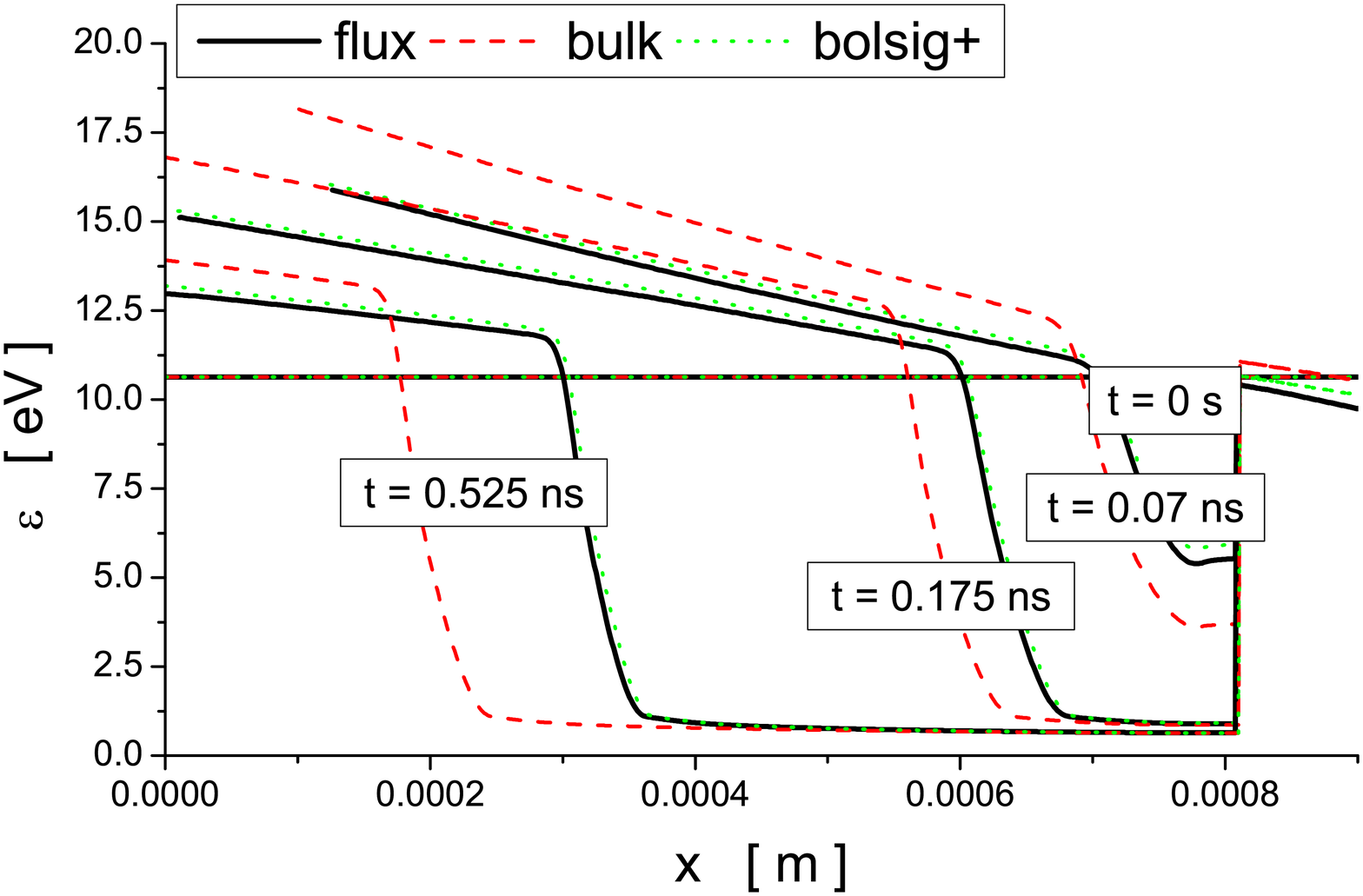}\\
\label{Fig3.2.5}
c) \includegraphics[scale=0.3]{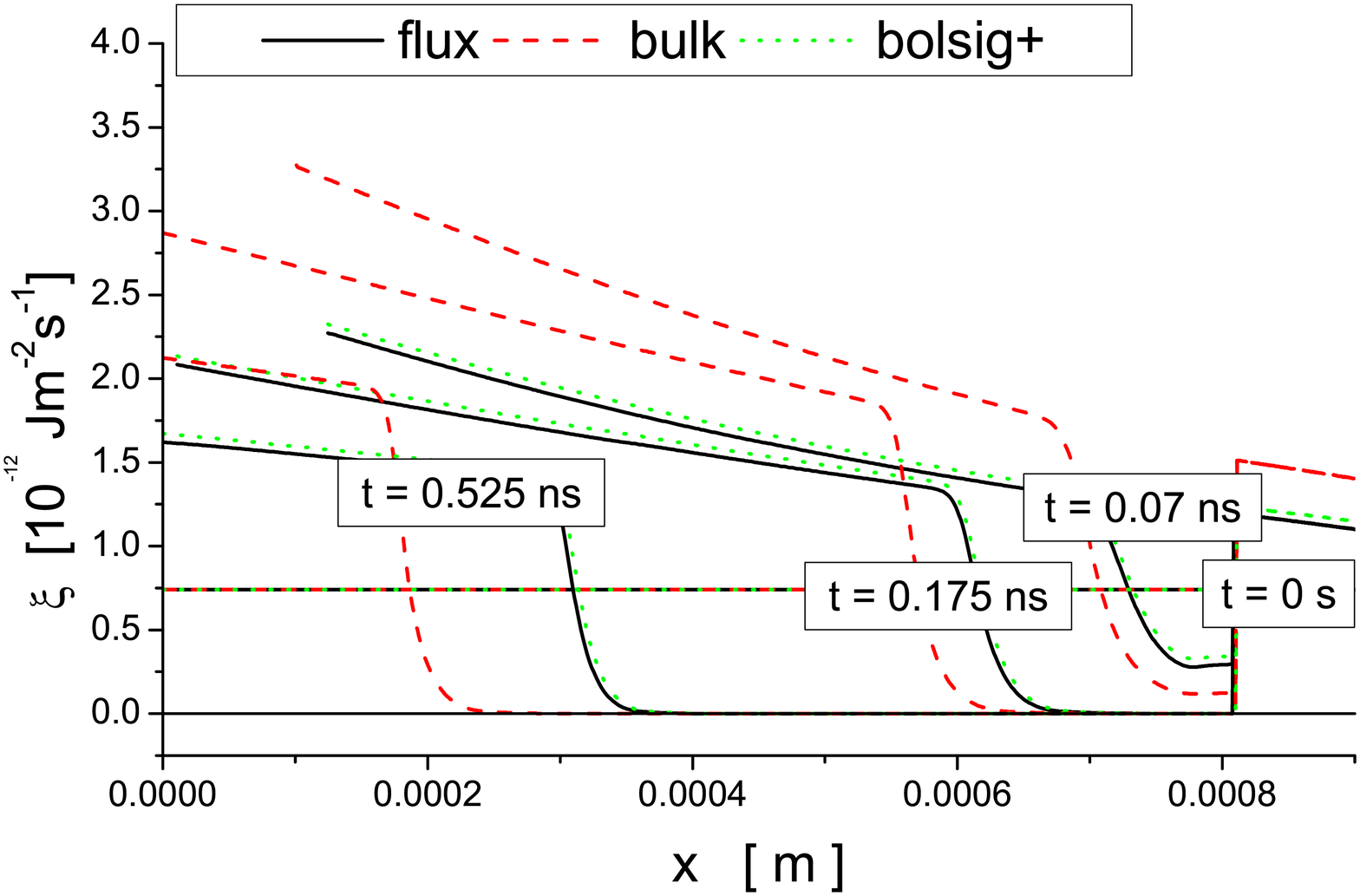}
\label{Fig3.2.7}
\caption{The same simulations and plots as in the previous figure, but now for (a) the average electron velocity $v$, (b) the average electron energy $\epsilon$, and (c) the average electron energy flux $\xi$. Again, velocity $v$ and energy flux $\xi$ here are negative quantities, and their absolute value is plotted.}
\label{fig7}
\end{figure}
\subsection{On the use of transport data in the high order fluid model}
\label{sec4.2}
Another important issue in streamer modeling concerns the choice of transport data in fluid models. The origin of, and the difference between bulk and flux transport coefficients in the context of streamer studies have been recently discussed in \cite{LiEH2010, LiEH2012, DujkoEWP2011}. In this section we consider the implications of the transport coefficient duality in the context of our high order fluid model for streamers. We also discuss to what extent the differences between transport coefficients obtained by the two term approximation for solving Boltzmann's equation and those obtained by a multi term theory affect the accuracy of streamer models.

Figure \ref{fig6} shows the temporal evolution of the electron and ion densities and of the electric field in a streamer front when the reduced electric field ahead of it is 590 Td. The calculations are performed for three different sets of input data as indicated in the figures. The inadequacy of using bulk data in the high order model is clearly evident (having in mind the very good agreement between the profiles obtained with the flux data and with the PIC/MC method, as shown in the previous section): the streamer velocity is higher and the electron/ion density in the streamer head and in the streamer channel is higher than with the flux transport data. For increasing electric field the discrepancies in the streamer velocity increase further, as illustrated in figure \ref{Fig3.1.4}. Similar trends have been observed in the preceding paper~\cite{PaperI} where planar fronts were modeled with the first order fluid model. In both models, streamers with bulk data are faster and the electron/ion density is higher in both the streamer head and the streamer channel. This follows directly from the fact that in N$_2$ the bulk mobility of the electrons is larger than the flux mobility. On the other hand, as already shown in section \ref{sec4.1} we observe an excellent agreement between our high order profiles obtained with the flux data and those predicted by a PIC/MC method.

Figure~\ref{fig7} displays the temporal evolution of the average electron velocity, of the average electron energy and of the averaged electron energy flux under the influence of electric field of 590 Td for different times as indicated on the graph. The inadequacy of using bulk data is again evident. In the early stage of the streamer development (when the electric field is not entirely screened), the average electron energy calculated using the bulk data is higher in all relevant streamer regions. When the transition process from an avalanche to a streamer is finished, we see a clear discrepancy between the average energies in the region ahead of the streamer front. As simulated planar fronts with bulk data are faster, the corresponding profiles of the average energy are shifted forward. However, for a fully developed streamer, there are no significant deviations between the average energies in the streamer channel. In this region the electric field is screened, and the average energies are slowly thermalizing although not fully relaxed as shown in previous section. The average electron velocity behaves exactly in the same manner as the average electron energy when comparing results with flux or bulk data. The only difference is associated with the fact that the average velocity of electrons is almost fully relaxed in the streamer channel regardless of the type of data used in calculations. The average electron energy flux shows similar behavior.

In earlier streamer investigations it has not been generally investigated to what extent the two term Boltzmann equation results for various transport coefficients affect the accuracy of the model predictions. The limitations of the two term approximation for solving Boltzmann's equation have been illustrated many times in past in the context of swarm studies \cite{WhiteRDNL2009, PetrovicDMMNSJSR2009, WhiteRSM2003, DujkoWPR2011}. Here the question arises of whether a similar conclusion can be drawn for streamers taking into account that the streamer development is a non-linear, non-stationary and non-hydrodynamic problem where space charge effects are important and where the electron energy varies from thermal values in the streamer channel up to a few tens of eV in the outer region of the streamer for the fields considered here. It is clear that the anisotropy of the velocity distribution function may be considerable in various streamer regions and hence we will compare profiles for different streamer properties using the two term and multi term results for transport data as an input in modeling.

Figure \ref{fig6}~(a) shows that the ionization level behind the streamer front calculated using the BOLSIG+ transport data is lower while the streamer velocity is the same. Surprisingly, the profiles of the electric field, the average electron energy and the average electron velocity obtained with the multi term flux data or with the BOLSIG+ data agree very well. This is a clear indication that internal errors of the two term approximation associated with the ionization rate can be directly used to understand the differences in the profiles of the electron and ion densities in the streamer channel (see figure 1 (c) in the preceding paper). However, this conclusion cannot be generalized to other gases or other conditions than those studied here. As pointed out recently by White and co-workers \cite{WhiteRDNL2009, WhiteRSM2003} one may never be sure about the accuracy of the two term approximation as various transport properties show different sensitivity with respect to this approximation in different energy ranges. For example, the inadequacy of the two term approximation has been recently demonstrated for CF$_4$ \cite{PetrovicDMMNSJSR2009} and CH$_4$ \cite{WhiteRSM2003}. In the preceding paper, it was shown that the velocity of a streamer in the first order fluid model is significantly affected by choosing either the two term or the multi term solution. Therefore, when high precision is required the best option is to use a multi term approach and/or a Monte Carlo simulation technique to calculate the input data for fluid models regardless of their order.

\subsection{Simulation of planar fronts with and without the energy flux term}
\label{sec4.3}

\begin{figure} [!htb]
\centering
\includegraphics[scale=0.35]{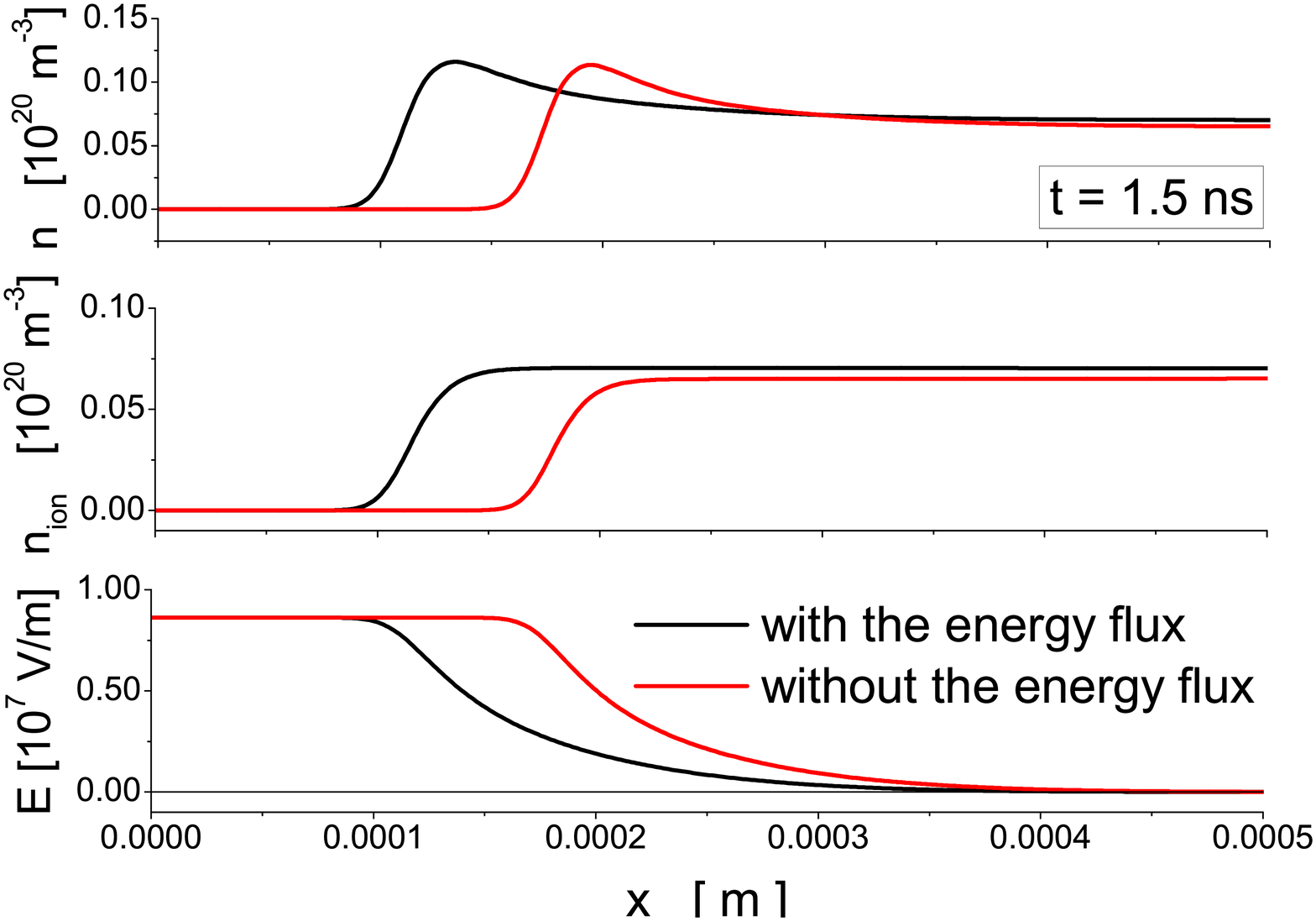}
\caption{Electron density, ion density and electric field in N$_2$ in an external electric field of 350 Td. The two lines indicate solutions of the high order fluid model with and without the energy flux term at the same instance of time. The initial condition is the same, and the flux transport data are used.}
\label{Fig3.3.1}
\end{figure}
\begin{figure} [!htb]
\centering
\includegraphics[scale=0.35]{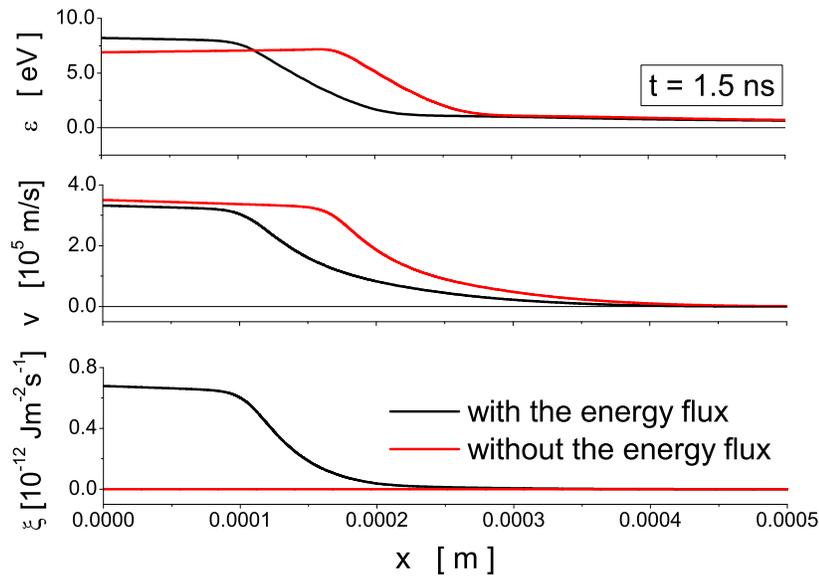}
\caption{The same instant and plot as in Fig.~\ref{Fig3.3.1}, but now average electron energy, negative average electron velocity and negative average electron energy flux are plotted.}
\label{Fig3.3.2}
\end{figure}

\begin{figure} [!htb]
\centering
a) \includegraphics[scale=0.35]{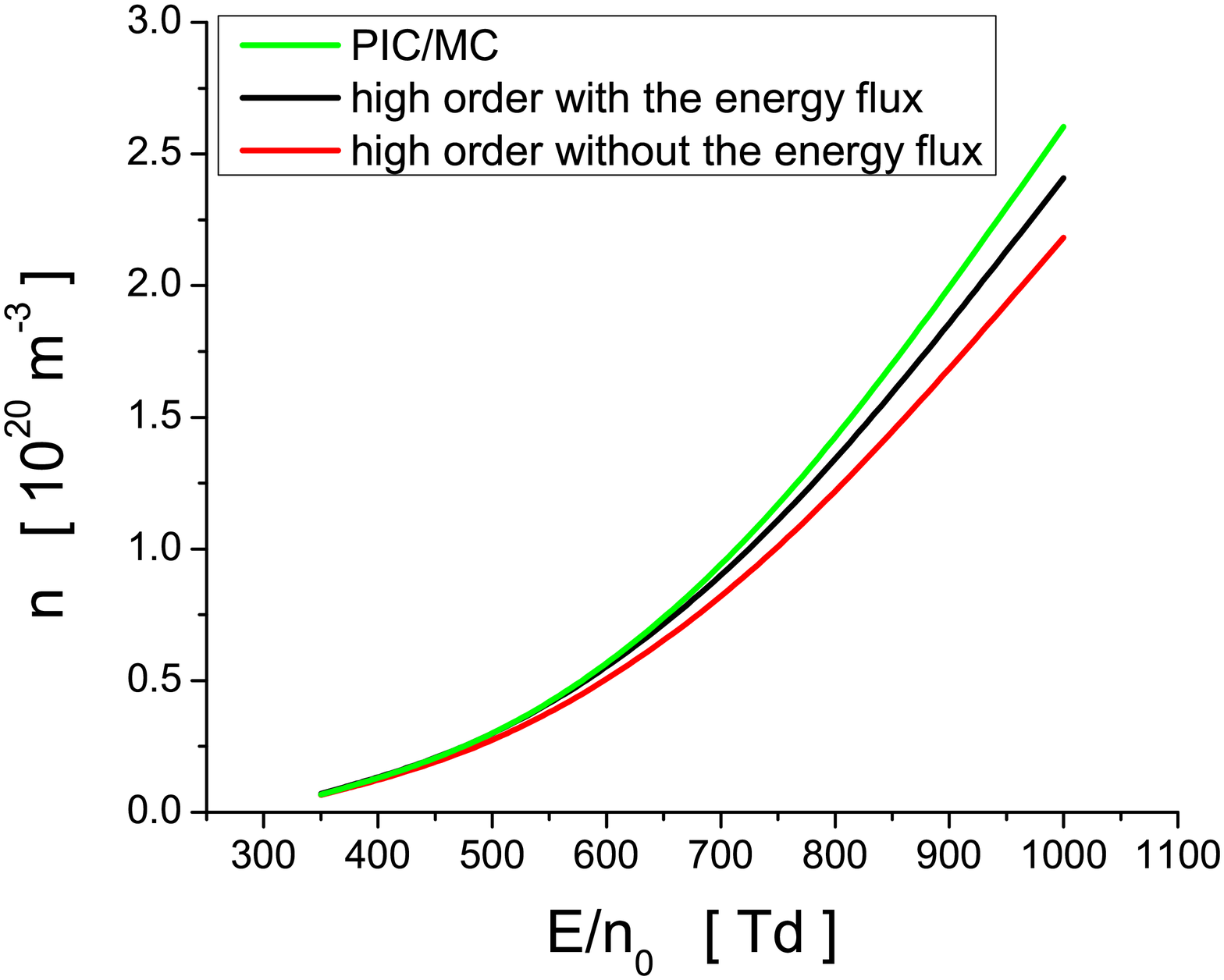}\\
b) \includegraphics[scale=0.35]{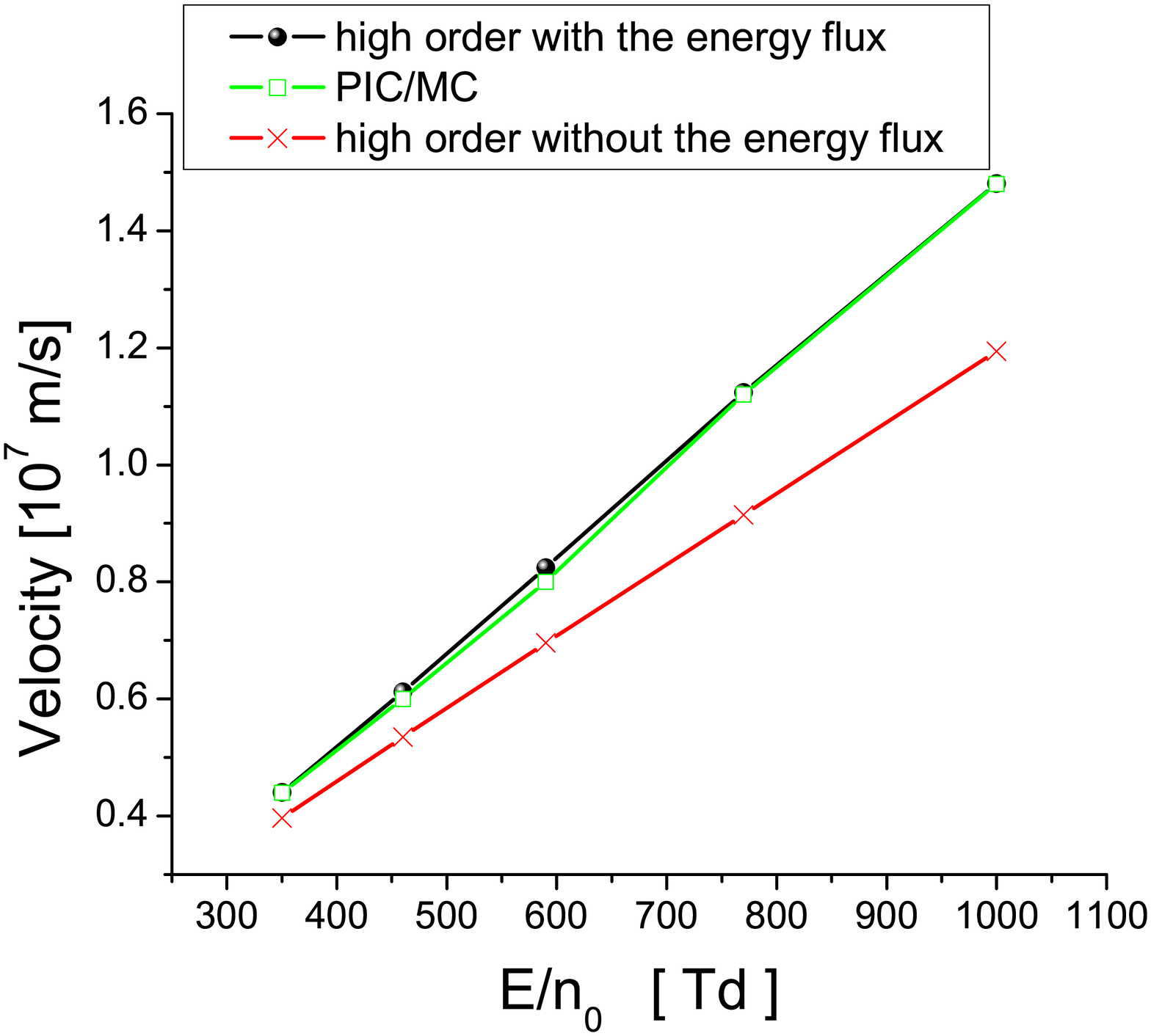}
\caption{(a) The ionization level behind the planar front, and (b) the absolute value of the front velocity as a function of the reduced electric field $E/n_0$ ahead of the front. Results are shown for the high order fluid model with or without the energy flux term, as well as for the PIC/MC model. Flux transport data are used as an input in the fluid models.}
\label{Fig3.3.6}
\end{figure}

In this section, we compare the results of the high order fluid model with and without the energy flux term (\ref{2.4}). In the model without the energy flux term the energy flux $\xi$ in the energy balance equation (\ref{2.3}) is set to zero, and therefore the energy flux balance equation (\ref{2.4}) does not need to be considered. This approach was already used in the case of streamer corona discharge dynamics \cite{KanzariYH1998, EichwaldDMYD2006} and for analysis of ionization wave dynamics in low-temperature plasma jets \cite{YousfiEMJ2012}. Figures \ref{Fig3.3.1} and \ref{Fig3.3.2} show the streamer properties for a reduced electric field $E/n_0$ of 350 Td at time $t=1.5$~ns. First, we see that the high order fluid model with the energy flux term gives higher electron and ion densities within the streamer channel than without the energy flux. For increasing $E/n_0$ these differences increase further, as shown in figure \ref{Fig3.3.6}(a). The electric field exhibits the typical streamer behavior; the observed differences between the high order fluid models with and without the energy flux term follow from the time delay needed for the space charge to become high enough and distort the externally applied spatially homogeneous electric field in these two models. This property is even more evident for the streamer velocity shown in figure \ref{Fig3.3.6}(b). Figures \ref{Fig3.3.1} and \ref{Fig3.3.2} show clearly that in our simulation conditions the streamer front propagates faster in the high order fluid model with the energy flux term.

Furthermore, the average electron energy and the average electron velocity ahead of the streamer depends clearly on the fact whether the electron energy flux is included or not. The average electron energy has a steep gradient between the outer streamer region and the streamer channel. In this spatially narrow region the energy decreases from the range between 10 and 20 eV to essentially thermal values in the streamer channel. It is clear that the correct treatment of the energy flux in the high order fluid model is critical for an accurate energy transport between the different streamer regions. In view of the close agreement between the present results with the energy flux term and those predicted by a PIC/MC method, established in the previous sections, we must conclude that the high order fluid model without the energy flux term is qualitatively right but quantitatively wrong under the conditions simulated.
\subsection{Simulation of planar fronts with and without the high order tensor}
\label{sec4.4}
In this section we investigate the influence of the high order tensor in the energy flux equation. In order to truncate the system after the energy flux equation, in the preceding paper this tensor was expressed by a product of two lower order tensors times a parameter $\beta$. In all previous sections, we used $\beta=1$ which is a natural guess. Here we investigate how the simulation results depend on $\beta$.
\begin{figure} [!htb]
\centering
\includegraphics[scale=0.30]{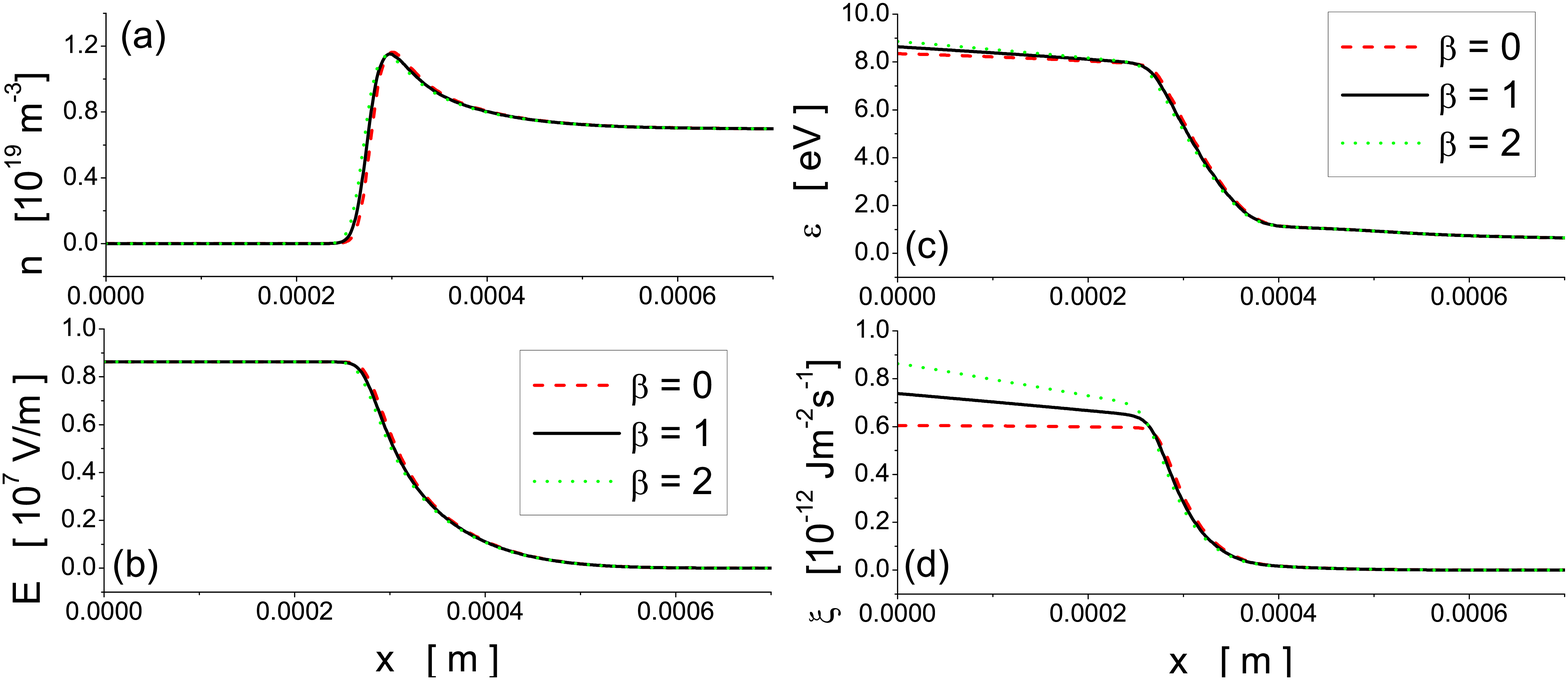}
\caption{Profiles of (a) electron density, (b) electric field, (c) average electron energy, and (d) average electron energy flux for $\beta=1$, $\beta=2$ and $\beta=3$. The results are shown for $E/n_0$ of 350 Td at time 1.5 ns.}
\label{Fig4.1.1}
\end{figure}

Figure \ref{Fig4.1.1} shows the profiles of the electron density, the electric field, the average electron energy and the average electron flux for a reduced electric field $E/n_0$ of 350 Td at time 1.5 ns. One can immediately see that increasing the parameter $\beta$ increases the velocity of the streamer very slightly. All other quantities remains essentially unaltered in the streamer channel indicating their weak sensitivity with respect to $\beta$. Only in the limit of the highest electric field considered here, the streamer properties in the channel respond slightly to the variation of the parameter $\beta$. Ahead of the streamer front, however, we observe variations in the streamer properties even for the reduced electric field of 350 Td. As expected, the most sensitive quantity with respect to variations in $\beta$ is the average electron energy flux.

In section \ref{sec2.1}, we have derived that our high order model is well posed if $\beta=0$ or $\beta\geq1$. We have solved our system for a range of values for $\beta$ and reduced electric fields $E/n_0$ and it was found that $\beta=1$ ensures the best agreement between profiles obtained by the high order fluid model and by the PIC/MC method. This validates the closure assumption associated with the contribution of high order tensors in the energy flux equation, as discussed in the preceding paper. Furthermore, if we analyze the profiles of various streamer properties for $\beta=0$, the simulation results agree very well with those derived by the PIC/MC method. The importance of this is twofold. First, the influence of the high order tensors in the energy flux equation is less than initially expected. Second, taking $\beta=0$ is a very good approximation, and it will significantly reduce the computation costs for the differential equations in 3D.

\section{Conclusion}
In the previous paper we have derived the equations and the transport and reaction parameters of a high order fluid model for streamer discharges. In the present paper we have briefly presented an accurate and efficient way to numerically solve the high order fluid model. These numerical schemes are then used to study the propagation of negative streamer fronts in N$_2$. Then we compared the high order fluid model and the first order fluid model with a PIC/MC model. While the first order model is very commonly used for streamer simulations, it can be clearly seen that the high order model is a much better approximation of the full particle dynamics. This concerns not only quantitative, but also qualitative differences. In particular we observe:\\
\noindent (a) The non-local effects in the profile of the average electron energy are present both in the PIC/MC-model and in the high order model, but are missing in the first order model due to the local field approximation. The slope of the average electron energy in the region ahead of the streamer is related to the spatial variation of the average electron energy in the avalanche phase of the streamer development. And the slope of the average electron energy in the streamer channel is striking and an inherent property of a streamer as well, as the electrons can not immediately relax to the low or vanishing field in the streamer interior.\\
\noindent (b) Bulk and flux transport coefficients for electrons can differ substantially in so-called non-conservative regions where ionization, attachment, detachment and recombination takes place. Here one needs to take care which coefficients to implement. Streamers in nitrogen with bulk data are faster and create a much higher electron density in the streamer channel. When the streamers in nitrogen are calculated with BOLSIG+ data, their velocity is not affected. However, in the streamer channel and in the streamer head we have found differences between the electron densities in our high order fluid model when we inserted either BOLSIG+ data or our flux data for the electron transport coefficients.\\
\noindent (c) When we conducted streamer simulations with our high order model, either including or excluding the energy flux, various streamer properties showed that the energy flux term cannot be neglected. Streamers without energy flux term are slower and create a lower electron density in the streamer channel. In the region ahead of the streamer, the average electron energy and the average electron velocity also change when the energy flux is neglected.\\
\noindent(d) The validity of the closure assumption associated with the treatment of high order terms in the energy flux equation is studied through the variation of the parameters used to parameterize the expression in which high order tensors are expressed in terms of lower order moments. We have pointed out that although the average electron energy and average electron energy flux in the outer region of the streamer are influenced by this parameter, the demand for a strict consideration of the high order tensor in the energy flux equation may be relaxed. This follows from the fact that it is almost impossible to notice the differences in the streamer velocity and electron density in the streamer channel for different parameters used to parameterize the high order terms. 

Though in the present paper, we have tested our model on planar fronts, the theory of the high order model has been developed in the previous paper without restrictions for the full three-dimensional setting. An important aspect to be addressed in the future is our closure assumption for the pressure tensor that we assumed to be isotropic. This restriction will be removed in future work.

Finally we would like to emphasize that though this study has concentrated on electron dynamics mainly in planar streamer fronts, the theory and the associated numerical solution of the system of differential equations are equally valid for ions up to the energy balance equation. Therefore we have derived a new modeling frame work that is widely applicable in reactive plasmas in full 3D.

\ack{It is a pleasure to acknowledge the helpful discussions with R.D White, W.H. Hundsdorfer, J. Teunissen and J. Bajars. AM and SD acknowledge support from STW-project 10751, SD acknowledges support from STW-project 10118, part of the Netherlands's Organization for Scientific Research (NWO). SD has also been supported by the MNTR, Serbia, under the contract number ON171037.}
\appendix
\section*{Appendix}
\setcounter{section}{1}

In this appendix we briefly describe the discretization of our system of differential equations. The finite volume method is used to discretize the system (\ref{2.7})-(\ref{2.12}) in space. The basic principle of the control volume method is to maintain the conservative properties of the system (\ref{2.7}) over every volume element. We introduce control volumes or cells $V_j$ as follows:
\begin{equation}
V_j:=\left[j\Delta x,(j+1)\Delta x\right],   \quad x_j:=\left(j+\frac{1}{2} \right)\Delta x, \quad j=0, 1,...,M-1,
\label{2.16}
\end{equation}
where $\Delta x=L/M$ is the spatial grid size while $L$ is the length of the simulation domain.

The numerical solution $\overline{\vc{u}}_j(t)$ has to be interpreted as an approximation of the average value of $\vc{u}(x,t)$ over the control volume $V_j$ at time $t$, i.e.
\begin{equation}
\overline{\vc{u}}_j(t)=\frac{1}{\Delta x}\int_{V_j}\vc{u}(x,t) \textrm{d}x.
\label{2.17}
\end{equation}

To approximate the spatial derivative in (\ref{2.7}) we use the second-order central difference discretization \cite{hund},
\begin{equation}
\frac{\partial \vc{u}(x,t)}{\partial x} = \frac{1}{2\Delta x} \left( \vc{u}(x+\Delta x,t) - \vc{u}(x-\Delta x,t) \right) +\mathcal{O}(\Delta x^2),
\label{2.18}
\end{equation}
which has already been used in previous first order fluid models and associated codes (see for example \cite{LiBEM2007}). Hence, the equation (\ref{2.7}) can be re-written in difference form:
\begin{equation}
\label{2.19}
\frac{\partial \overline{\vc{u}}_j(t)}{\partial t} + \frac{1}{2\Delta x}\vc{ A}(\overline{\vc{u}}_j(t)) (\overline{\vc{u}}_{j+1}(t)-\overline{\vc{u}}_{j-1}(t))=\vc{ F} (\overline{\vc{u}}_j(t)),
\label{2.20}
\end{equation}
where $i= 1,2,...., M-2$.

{As discussed in the main text, homogeneous Neumann or Dirichlet boundary conditions are imposed:
At the left boundary, $i = 0$, we impose a homogeneous Neumann boundary condition
\begin{equation}
\frac{\partial \vc{u}(0,t)}{\partial x} = 0 \quad \textrm{or}\quad \overline{\vc{u}}_{-1}(t) = \overline{\vc{u}}_{0}(t)\,.
\label{2.21}
\end{equation}
In this manner, we provide the value $\overline{\vc{u}}_{-1}(t) = \vc{u}(-\frac{1}{2}\Delta x,t)$ at $i=-1$ which exceeds the computational domain. At the right boundary, $i = M-1$, we impose a homogeneous Dirichlet boundary condition for the electron number density, while for the other components of $\vc u$ we impose homogeneous Neumann boundary conditions to obtain a value for $\overline{\vc{u}}_{M}(t) = \vc{u}(L+\frac{1}{2}\Delta x,t)$.

To approximate the electric potential $\varphi$ (\ref{2.6}) we use the same strategy as for the primitive variables $\mathbf{u}$, i.e. $\varphi$ is averaged over control volumes. Hence, the electric field $E(x,t)$ is discretized on the edges of the control volume, denoted by $E(x_{j-\frac{1}{2}},t)$. But since we consider a 1D case, the electric field follows directly from the charge densities by integrating equation (\ref{2.12}) along the $x$-direction:
\begin{equation}
E(x,t) = E(0,t) + \frac{e}{\epsilon_0}\left(\int^{x}_0 (n_{ion}(x',t)-n(x',t)\right)\textrm{d} x'.
\label{2.22}
\end{equation}
We then can calculate $E(x_j,t)$ by:
\begin{equation}
E(x_j,t) = \frac{1}{2}(E(x_{j-\frac{1}{2}},t) + E(x_{j+\frac{1}{2}},t) ).
\label{2.24}
\end{equation}

Now we are going to consider the ODE system (\ref{2.19}). To approximate the time derivatives we use classical RK4 (Rounge-Kutta 4) time-integration scheme \cite{hund}, which is a fourth order method. This is an explicit method which always has a bounded stability domain. In our case the stability condition has the following form:
\begin{equation}
\mathop{\mbox{max}}_{1\le i \le 4} | \lambda_i \Delta t /2\Delta x | \leq C,
\label{2.25}
\end{equation}
or by taking into account $\beta\geq \sqrt{\beta+\sqrt{\beta(\beta-1)}}$, when $\beta \geq 1$, we have:
\begin{equation}
\beta \frac{\Delta t }{2\Delta x }\sqrt{\frac{2\max \varepsilon}{3m}}\leq C.
\label{2.26}
\end{equation}
This condition is called CFL stability criterion, where $C$ depends on the particular time-integration method and space discretization. In our simulations, $C$ is set to 0.1.



\section*{References}
\begin{thebibliography}{99}

\bibitem{Pasko2006} M. Füllekrug Martin, E.A. Mareev and M.J. Rycroft (Editors), Proceedings of the NATO Advanced Study Institute on {\it Sprites, Elves and Intense Lightning Discharges}, Corte, Corsica, France, 24-31 July 2004, (Dordrecht, Springer) (2006) 
\bibitem{EbertS2008} U. Ebert and D.D. Sentman, {\it J. Phys. D: Appl. Phys.} {\bf 41} 230301 (2008) 
\bibitem{EbertNLLBV2010} U. Ebert, S. Nijdam, C. Li, A. Luque, T.M.P. Briels and E.M. van Veldhuizen, {\it J. Geophys. Res.} {\bf 115} A00E43 (2010)
\bibitem{LuqueE2010} A. Luque and U. Ebert, {\it Geophys. Res. Lett.} {\bf 37} L06806 (2010)
\bibitem{Flesch2006} P. Flesch, {\it Light and light sources, High-intensity discharge lamps} (Berlin: Springer) (2006)
\bibitem{ListerLL2004} G.G. Lister, J.E. Lawler, W.P. Lapatovich and V.A. Godyak, {\it Rev. Mod. Phys.}{\bf 76} 541 (2004)
\bibitem{SobotaMVDH2010} A. Sobota, F. Manders, E.M. van Veldhuizen, J. van Dijk and M. Haverlag, {\it IEEE Trans. Plasma Sci.} {\bf 38} 2289 (2010)
\bibitem{Veldhuizen2000} E.M. van Veldhuizen, {\it Electrical Discharges for Environmental Purposes: Fundamentals and Applications} (Huntington, NY: Nova Science) (2000)
\bibitem{vanHeeschWP2008} E. van Heesch, G. Winands and A. Pemen, {\it J. Phys. D: Appl. Phys.} {\bf 41} 234015 (2008)
\bibitem{Naidis2010} G.V. Naidis, {\it J. Phys. D: Appl. Phys.} {\bf 43} 402001 (2010)
\bibitem{Naidis2011} G. Naidis, {\it J. Phys. D: Appl. Phys.} {\bf 44} 215203 (2011)
\bibitem{LuLP2012} X.P. Lu, M. Laroussi and V. Puech, {\it Plasma Sources Sci. and Technol.} {\bf 21} 034005 (2013)
\bibitem{YousfiEMJ2012} M. Yousfi, O. Eichwald, N. Merbahi and N. Jomaa, {\it Plasma Sources Sci. Technol.} {\bf 21} 045003 (2012) 
\bibitem{BoeufYP2013} J-P. Boeuf, L.L. Yang, L.C. Pitchford, {\it J. Phys. D: Appl. Phys.} {\it 46} 015201 (2013)
\bibitem{Starikovskaia2006} A. Starikovskaia, {\it J. Phys. D: Appl. Phys.} {\bf 39} R265 (2006)
\bibitem{StarikovskiyA2013} A. Starikovskiy and N. Aleksandrov, {\it Progress in Energy and Combustion Science} {\bf 39} 61 (2013) 
\bibitem{EbertMBHMRV2006} U. Ebert, C. Montijn, T.M.P. Briels, W. Hundsdorfer, B. Meulenbroek, A. Rocco, E.M. van Veldhuizen, {\it Plasma Sources Sci. and Technol.} {\bf 15} S118 (2006)
\bibitem{LiBEM2007} C. Li, W.J.M. Brok, U. Ebert and J.J.A.M. van der Mullen, {\it J. Appl. Phys.} {\bf 101} 123305 (2007)
\bibitem{LiEH2010} C. Li, U. Ebert and W. Hundsdorfer, {\it J. Comp. Phys.} {\bf 229} 200 (2010)
\bibitem{LiEH2012} C. Li, U. Ebert and W. Hundsdorfer, {\it J. Comp. Phys.} {\bf 231} 1020 (2012)
\bibitem{LuqueE2012} A. Luque and U. Ebert, {\it J. Comp. Phys.} {\bf 231} 904 (2012)
\bibitem{LiTNHE2012} C. Li, J. Teunissen, M. Nool, Willem Hundsdorfer and U. Ebert, {\it Plasma Sources Sci. Technol.} {\bf 21} 055019 (2012)
\bibitem{LiEBH2008} C. Li, U. Ebert, W.J.M. Brok and W. Hundsdorfer, {\it J. Phys. D: Appl. Phys.} {\bf 41} 032005 (2008)
\bibitem{LiEH2009} C. Li, U. Ebert and W. Hundsdorfer, {\it J. Phys. D: Appl. Phys.} {\bf 42} 2020003 (2009)
\bibitem{BayleC1985} P. Bayle and B. Cornebois, {\it Phys. Rev. A} {\bf 31} 1046 (1985)
\bibitem{GuoW1993} J.M. Guo and C.H.J. Wu, {\it IEEE Trans. Plasma Sci.} {\bf 21} 684 (1993)
\bibitem{Naidis1997} G.V. Naidis, {\it Tech. Phys. Lett.} {\bf 23} 493 (1997)
\bibitem{KanzariYH1998} Z. Kanzari, M. Yousfi and A. Hamani, {\it J. Appl. Phys.} {\bf 84} 4161 (1998)
\bibitem{EichwaldDMYD2006} O. Eichwald, O. Ducasse, N. Merbahi, M. Yousfi and D. Dubois, {\it J. Phys. D: Appl. Phys.} {\bf 39} 99 (2006)
\bibitem{KorenEGGKK2012} B. Koren, U. Ebert, T. Gombosi, H. Guillard, R. Keppens, D. Knoll, {\it J. Comput. Phys.} {\bf 231} 717 (2012) 
\bibitem{PaperI} S. Dujko, A.H. Markosyan, R.D. White and U. Ebert, {\it High order fluid model for streamer discharges. I. Derivation of model and transport data}, manuscript submitted together with the present manuscript
\bibitem{Robson1986} R.E. Robson, {\it J. Chem. Phys.} {\bf 85} 4486 (1986)
\bibitem{VrhovacP1996} S.B. Vrhovac and Z.Lj. Petrovi´c, {\it Phys. Rev. E} {\bf 53} 4012 (1996)
\bibitem{WhiteRDNL2009} R.D. White, R.E. Robson, S. Dujko, P. Nicoletopoulos and B.Li, {\it J. Phys. D: Appl. Phys.} {\bf 42} 194001 (2009)
\bibitem{LuqueGV2011} A. Luque and F.J. Gordillo-Vázquez, {\it Nature Geosci.} {\bf 5} 22 (2011)
\bibitem{FlittiP2009} A. Flitti and S. Pancheshnyi, {\it Eur. Phys. J. Appl. Phys.} {\bf 45} 1001 (2009)
\bibitem{Pancheshnyi2005} S. Pancheshnyi, {\it Plasma Sources Sci. Technol.} {\bf 14} 645 (2005)
\bibitem{Popov2001} N.A. Popov, {\it Plasma Phys. Reports} {\bf 27} 886 (2001)
\bibitem{Popov2003} N.A. Popov, {\it Plasma Phys. Reports} {\bf 29} 695 (2003)
\bibitem{SentmanSN2009} D.D. Sentman and H.C. Stenbaek-Nielsen, {\it Plasma Sources Sci. Technol.} {\bf 18} 034012 (2009)
\bibitem{Leveq1} R.J. LeVeque, {\it Finite Difference Methods for Ordinary and Partial Differential Equations: Steady-State and Time-Dependent Problems} (SIAM, Philadelphia) (2007)
\bibitem{Leveq2} R.J. LeVeque, {\it Finite Volume Methods for Hyperbolic Problems} (Cambridge University Press, London) (2002)
\bibitem{MortonM} K.W. Morton and D.F. Mayers, {\it Numerical Solution of Partial Differential Equations} (Cambridge University Press, London) (2005)
\bibitem{DujkoWRP2012} S. Dujko, R.D. White, Z.M. Raspopovi\'{c} and Z.Lj. Petrovi\'{c}, {\it Nucl. Intr. Meth. B} {\bf 279} 84 (2012)
\bibitem{PetrovicRDM2002} Z.Lj. Petrovi\'{c}, Z.M. Raspopovi\'{c}, S. Dujko and T. Makabe, {\it Appl. Surf. Sci.} {\bf 192} 1 (2002)
\bibitem{WhiteNR2002} R.D. White, K.F. Ness and R.E. Robson {\it Appl. Surf. Sci.} {\bf 192} 26 (2002)
\bibitem{DujkoEWP2011} S. Dujko, U. Ebert, R.D. White and Z.Lj. Petrovi\'{c}, {\it Jpn. J. Appl. Phys.} {\bf 50} 08JC01 (2011)
\bibitem{NessR1986} K.F. Ness and R.E. Robson, {\it Phys. Rev. A} {\bf 34} 2185 (1986)
\bibitem{PetrovicDMMNSJSR2009} Z.Lj. Petrovi\'{c}, S. Dujko, D. Mari\'{c}, G. Malovi\'{c}, \v{Z} Nikitovi\'{c}, O. \v{S}a\v{s}i\'{c}, J. Jovanovi\'{c}, V. Stojanovi\'{c} and M. Radmilovi\'{c}-Radjenovic, {\it J. Phys. D: Appl. Phys.} {\bf 31} 194002 (2009)
\bibitem{WhiteRSM2003} R.D. White, R.E. Robson, B. Schmidt and M.A. Morrison, {\it J. Phys. D: Appl. Phys.} {\bf 36} 3125 (2003)
\bibitem{DujkoWPR2011} S. Dujko, R.D. White, Z.Lj. Petrovi'{c} and R.E. Robson, {\it Plasma Sources Sci. Technol.} {\bf 20} 024013 (2011)
\bibitem{hund} W. Hundsdorfer, J.G. Verwer, {\it Numerical Solution of Time-Dependent Advection-Diffusion-Reaction Equations} (vol. 33 of {\it Series in Comp. Math.} Springer, Berlin) (2003)
\bibitem{Aleksandrov1996} N. Aleksandrov, I. Kochetov, {\it J. Phys. D: Appl. Phys.} {\bf 29} 1476 (1996)
\bibitem{Naidis1997b} G.V. Naidis, {\it Pis'ma Zh. Tekh. Fiz.} {\bf 23} 89 (1997)

\end{thebibliography}
\end{document}